\documentclass[prb, aps, twocolumn,superscriptaddress,preprintnumbers]{revtex4-2}
\usepackage[]{amssymb}

\usepackage{amsmath}
\usepackage{makeidx}
\usepackage{amsfonts}
\usepackage[ansinew]{inputenc}
\usepackage[usenames,dvipsnames]{pstricks}
\usepackage{subfigure}
\usepackage{epsfig}
\usepackage{graphicx}
\usepackage{dcolumn}
\usepackage{bm}
\usepackage[colorlinks,hyperindex]{hyperref}
\usepackage{epstopdf}	
\usepackage{float}

\begin{document}
	
\title[Modelling the acoustic response]{Analytical model of the acoustic response of nanogranular films adhering on a substrate}

\author{Gianluca Rizzi}
\affiliation{\footnotesize GEOMAS, INSA-Lyon, Universit\'{e} de Lyon, 20 avenue Albert Einstein, 69621, Villeurbanne cedex, France}
\author{Giulio Benetti}
\email{giulio.benetti@aovr.veneto.it}
\affiliation{\footnotesize Department of Pathology and Diagnostics-Medical Physics Unit, University Hospital of Verona, P.le Stefani 1, 37126 Verona, Italy}
\author{Claudio Giannetti}
\affiliation{\footnotesize Interdisciplinary Laboratories for Advanced Materials Physics (i-LAMP) and Dipartimento di Matematica e Fisica, Universit{\`a} Cattolica del Sacro Cuore, Via Musei 41, 25121 Brescia, Italy}
\author{Luca Gavioli}
\affiliation{\footnotesize Interdisciplinary Laboratories for Advanced Materials Physics (i-LAMP) and Dipartimento di Matematica e Fisica, Universit{\`a} Cattolica del Sacro Cuore, Via Musei 41, 25121 Brescia, Italy}
\affiliation{\footnotesize FemtoNanoOptics group, Universit\'{e} de Lyon, CNRS, Universit\'{e} Claude Bernard Lyon 1, Institut Lumi\`{e}re Mati\`{e}re, F-69622 Villeurbanne, France}
\author{Francesco Banfi}
\affiliation{\footnotesize FemtoNanoOptics group, Universit\'{e} de Lyon, CNRS, Universit\'{e} Claude Bernard Lyon 1, Institut Lumi\`{e}re Mati\`{e}re, F-69622 Villeurbanne, France}


\begin{abstract}
A 1D mechanical model for nanogranular films, based on a structural interface, is here presented.
The analytical dispersion relation for the frequency and lifetimes of the acoustics breathing modes is obtained in terms of the interface layer thickness and porosity.
The model is successfully benchmarked both against 3D Finite Element Method simulations and experimental photoacoustic data on a paradigmatic system available from the literature. A simpler 1D model, based on an homogenized interface, is also presented and its limitations and pitfalls discussed at the light of the more sophisticated pillar model. The pillar model captures the relevant physics responsible for acoustic dissipation at a disordered interface. Furthermore, the present findings furnish to the experimentalist an easy-to-adopt, benchmarked analytical tool to extract the interface layer physical parameters upon fitting of the acoustic data. The model is scale invariant and may be deployed, other than the case of granular materials, where a patched interface is involved. 
\end{abstract}

\maketitle

\section{Introduction}
\begin{figure*}
	\centering
	\includegraphics[width=2\columnwidth]{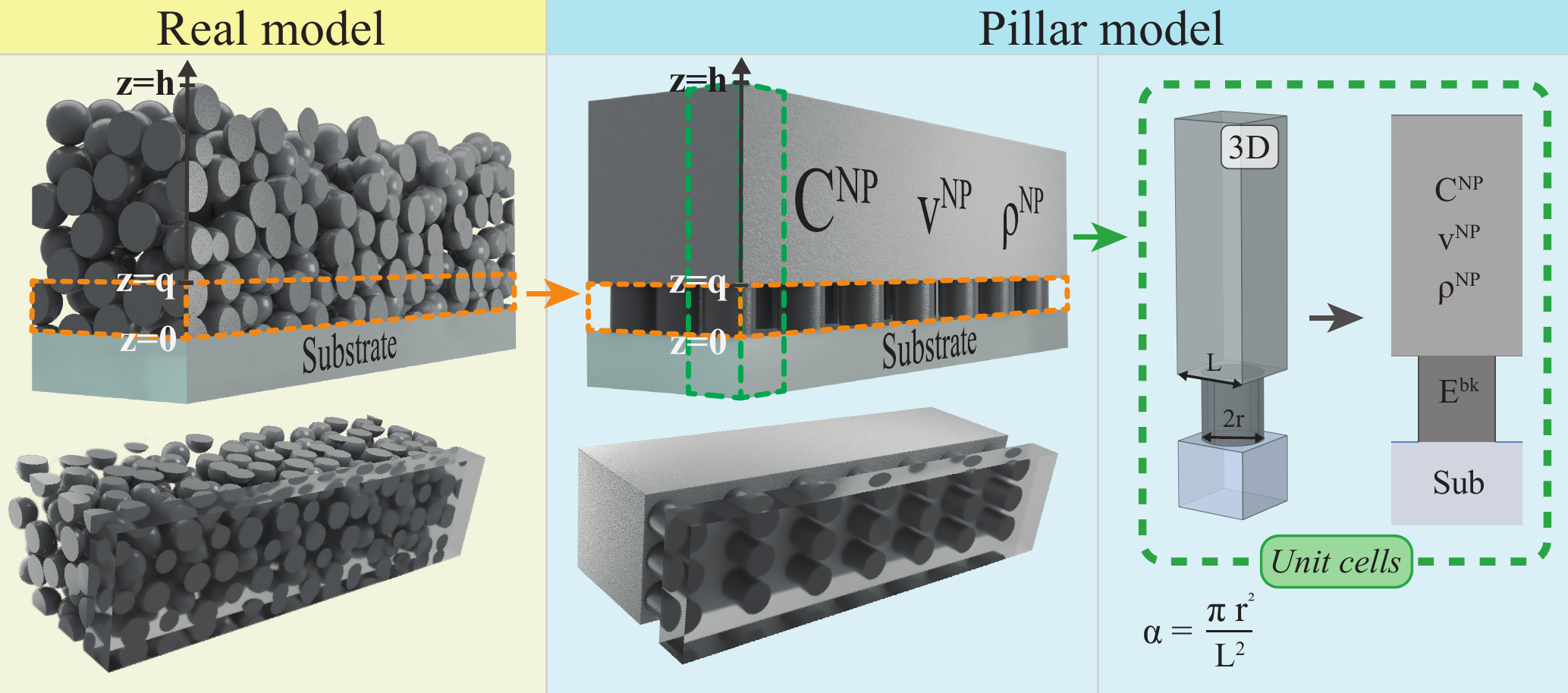}
	\caption{
		Left: 3D nanoparticles thin film of thickness $h$ adhered on a semi-infinite substrate. The bottom view, as seen looking across the substrate, highlights the `patched' interface.
		Centre: 3D Pillar model: \textit{effective} NP layer ($q<z<h$); pillars layer ($0<z<q$); semi-infinite substrate ($z<0$).
		The NP layer \textit{effective} density and stiffness tensor are $\rho^{NP}$ and $C^{NP}$, respectively. The pillars layer density, $\rho^{bk}$, and Young modulus, $E^{bk}$, are the same as the ones of the material of which the NPs are made (see text). The bottom view, as seen looking across the substrate, highlights the similarity with the `patched' interface of the real case. Right: reduction of the periodic 3D Pillar model to a single 3D unit cell of base size $L\times L$. The pillar layer filling fraction, $\alpha$, is defined as the ratio of the pillar cross-sectional area to that of the unit cell, irrespective of the geometry of the pillar cross section. The image is for illustrative purposes.
	}
	\label{fig:geometry_scheme}
\end{figure*}
Nanogranular ultrathin films are at the forefront of a wide range of technological applications \cite{stark2015industrial} ranging from nanomedicine \cite{benetti2019tailored}, sensing \cite{villa2019soft,benetti2018photoacoustic,huang2017fast} to electronics \cite{nasiri2016tunable,minnai2017facile,caruso2016high,santaniello2020additive,mirigliano2019non,tarantino2020,mirigliano2021}.
Accessing their mechanical properties, both within the film's bulk and at the interface region in contact with the supporting substrate, is among the most urgent issues in view of any device development.
In this context photoacoustic nanometrology plays a key role.
For instance, the bulk properties of periodic nanogranular thin films have been explored across a variety of configurations \cite{tournat2010acoustics,boechler2017dynamics} ranging from 1D \cite{allein2016tunable}, 2D \cite{hiraiwa2016complex,vega2017vibrational,wallen2015dynamics,rizzi2020exploring,graczykowski2020,ghanem2020,babacic2020} to 3D \cite{abi2019longitudinal,merkel2010dispersion} arrangements.
Recently, the development of table-top UV laser sources allowed generating surface acoustic waves with periodicity in the 10 nm range \cite{siemens2009high}, hence opening to mechanical nanometrology \cite{nardi2015impulsively} of periodic granular thin films of thicknesses down to few nanometers \cite{abad2020nondestructive,frazer2020full}. Photoacoustics investigations of the \textit{bulk} properties of \textit{non-periodic} nanogranular films have also been performed in several contexts over granularities ranging from few nm \cite{peli2016mechanical,benetti2017bottom,benetti2018photoacoustic}, to hundreds of nm \cite{ayouch2012elasticity,girard2018contact} up to the micron scale \cite{hiraiwa2017acoustic}. As for \textit{interface properties}, photoacoustic investigations mainly focused on \textit{homogeneous} thin films \cite{tas1998picosecond,dehoux2009nanoscale,dehoux2010picosecond,hettich2011modification,ma2015comprehensive,hoogeboom2016nondestructive,hettich2016viscoelastic,grossmann2017characterization,greener2019high,zhang2020unraveling,clemens2020}, nanogranular thin film interfaces remaining relatively unexplored.
The difficulty is to address `patched' interfaces as the one emerging between an aperiodic granular film and the adhering substrate, disorder being the critical aspect \cite{peli2016mechanical}.
Acoustic attenuation times for such an interface are hard to conceive in analytical terms, calling for full 3D Finite Element Method (FEM) simulations and casting the acoustic wave problem at the interface in scattering terms.
These approaches, whenever applicable, do not shed much light on the underlying physics and are hardly implementable to fit photoacoustic data due to computational costs. Furthermore, implementation of full 3D models requires knowledge of the detailed film morphology at the interface which, for the case of aperiodic granular materials, is unknown or very difficult to achieve \cite{benetti2017bottom}.  
Therefore, easy-to-adopt mechanical models are necessary to interpret photoacoustics data, retrieving the interface
physical properties and ultimately unveiling the relevant physics ruling the acoustic to structure relation in materials with disordered interfaces.
From a general view point, the situation here addressed is complementary to that of acoustic damping from a single nano-object to its supporting substrate \cite{hartland2011optical,devkota2019making}. For the latter, the experimental is challenging whereas the modelling is rather straight forward since it relies on a thorough system's knowledge \cite{maioli2018mechanical,devkota2018measurement}. On the contrary, in the present case the experimental is relatively simple \cite{peli2016mechanical}, the modelling though is the delicate and yet unsolved issue. This is ascribable to the disordered, hence intrinsically undetermined, interface.

A 1D mechanical model for nanogranular thin films adhered on a flat substrate is here proposed. The model, addressed as pillar model, is based on a structural interface \cite{bertoldi2007structural}, meaning that a true structure is introduced to mimic the transition region between the NP's film bulk and the underlying substrate.
Extrinsic attenuation, i.e. acoustic radiation to the substrate, is assumed to prevail over intrinsic attenuation which is not accounted for.
The analytical dispersion relation for the frequencies and lifetimes of the ultrathin film's acoustic breathing modes, i.e. the ones commonly excited in photoacoustic experiments, is obtained in terms of the interface layer physical parameters: interface porosity and layer thickness. 
The model is successfully benchmarked both against a full 3D FEM model and against experimental photoacoustic data available from the literature on a paradigmatic model system, in which knowledge of mechanical properties at the interface is a key asset in a variety of applications \cite{benetti2018photoacoustic,torrisi2019ag,benetti2020antimicrobial}. A simpler 1D model, addressed as Effective Medium Approximation model (EMA) and based on an homogenized interface layer, is also provided together with its dispersion relation. Its limits of validity, restrained to small porosities, are discusses at the light of the pillar model. Assuming the granular film made of nanoparticles (NP), the present theoretical scheme is here tested for the case of NP radiuses smaller than the film thicknesses and inferior to the excited breathing modes wavelength.

The pillar model rationalises the acoustic to structure relation in materials affected by disordered interfaces. The physics is here shown to be ruled by the integral of the stresses exchanged across the interfaces rather than their detailed distribution. The pillar model, on one side, furnishes to the experimentalist an experimentally-benchmarked, easy-to-adopt analytical tool to extract the interface layer physical parameters upon fitting of the acoustic data. On the other side, upon previous knowledge of the interfacial layer parameters, the model allows retrieving the breathing modes frequencies and lifetimes of a nanogranular coating adhering on a substrate. All these aspects bear both a fundamental and applicative interest across a wide range of fields ranging from condensed matter, material science to device physics.

\section{The pillar model}
\begin{figure*}
	\centering
	\begin{minipage}[b]{\columnwidth}
		\includegraphics[width=1.03\columnwidth]{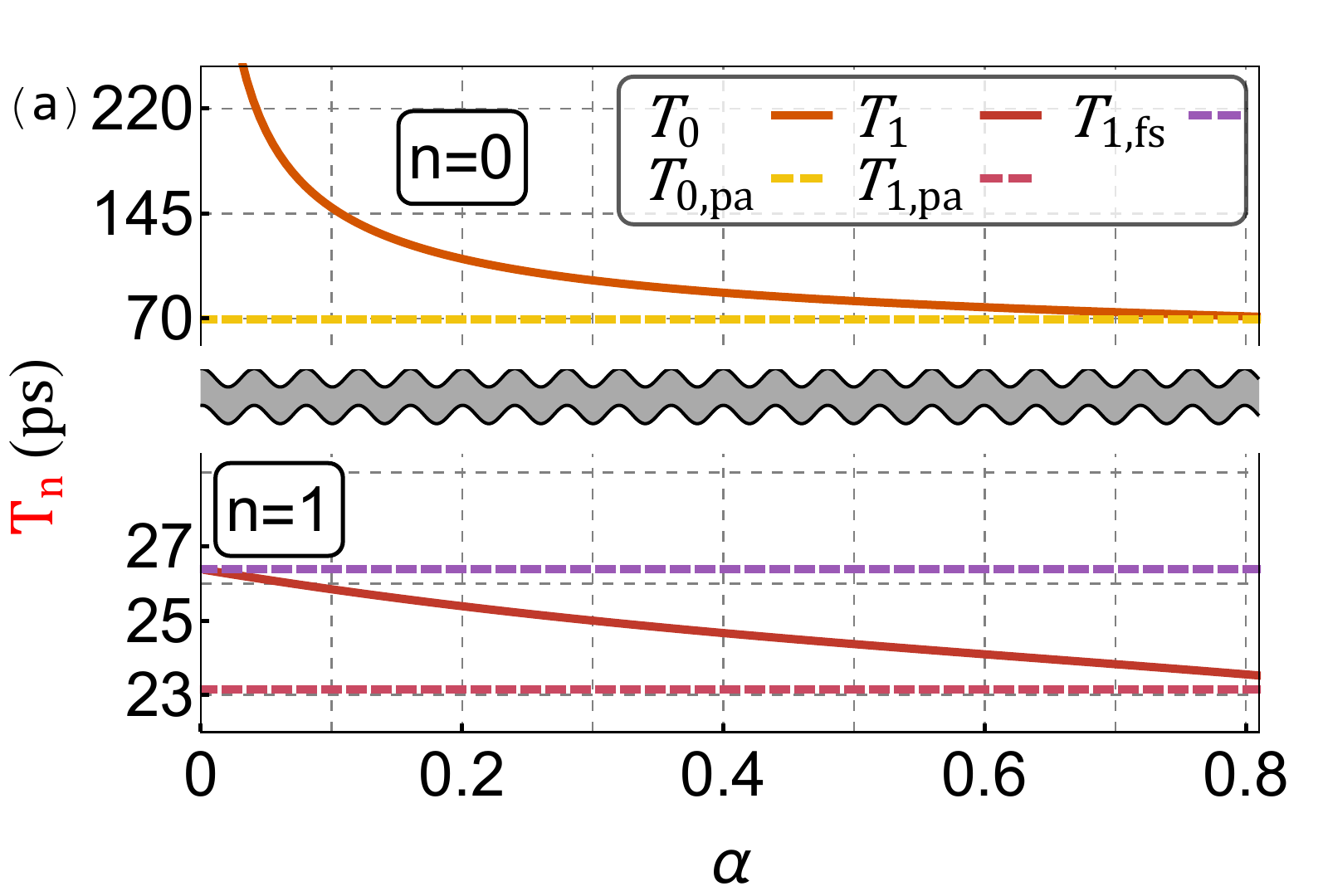}
	\end{minipage}
	\hfill
	\begin{minipage}[b]{\columnwidth}
		\includegraphics[width=\columnwidth]{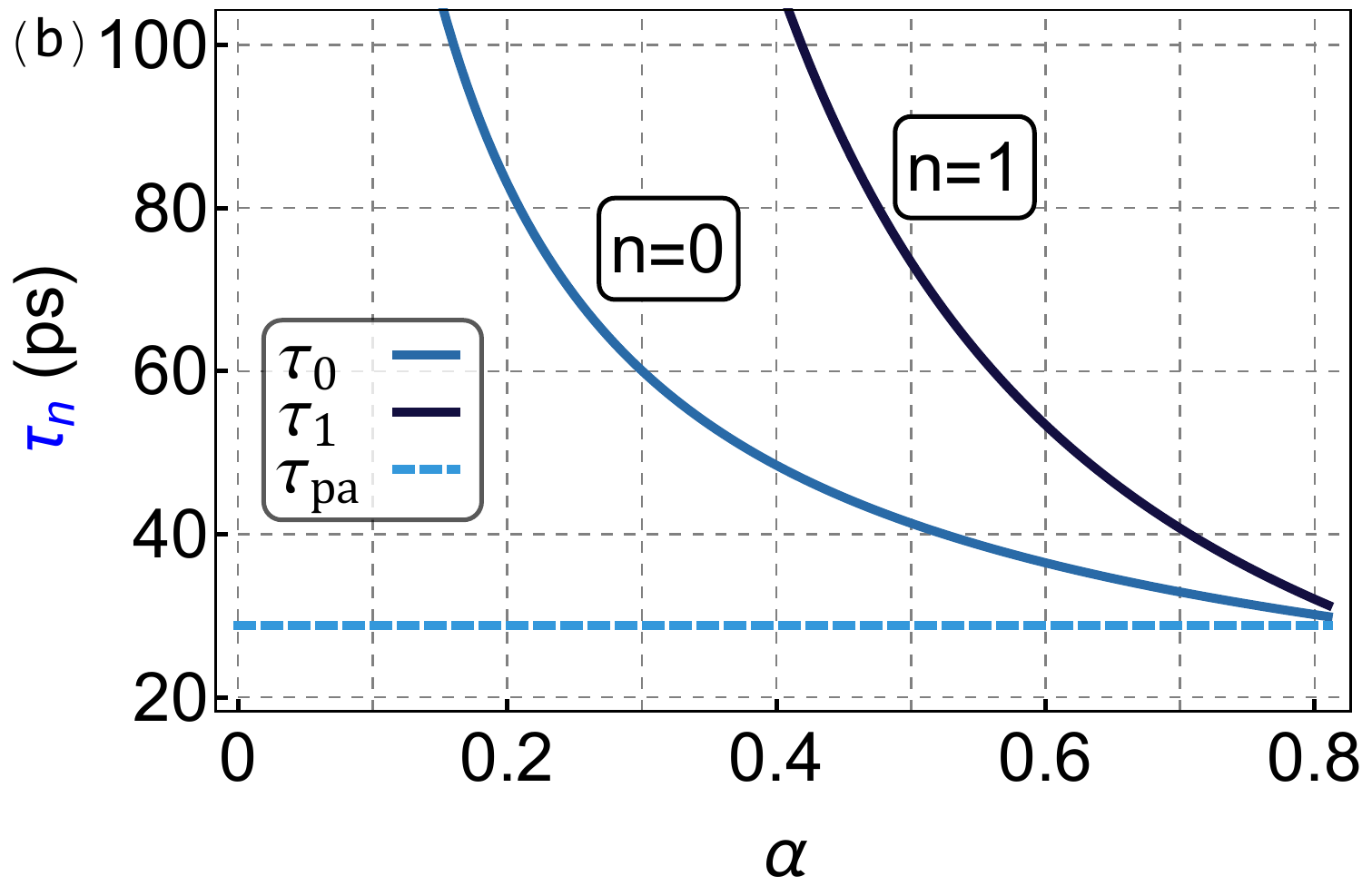}
	\end{minipage}
	\caption{(a) period $T_n$ and (b) decay time $\tau_n$ versus $\alpha$ for the Ag nanogranular film (see text) with $q$=12 nm and $h$=50 nm. The first two modes $n=\{0,1\}$ are addressed. Pillar model (full lines) and limit cases (dashed lines) obtained for the free standing (subscript '$fs$') and perfect adhesion (subscript '$pa$') scenarios respectively. The y-axis in the graph of panel (a) is broken for sake of graphical clarity. The scales above and below the brake are different for ease of representation. The $fs$ scenario yields, for mode $n$=0, an infinite period (corresponding to a film translation), hence it is not reported in panel (a). The $fs$ scenario yields an infinite decay time, hence it is not reported in panel (b).}
	\label{fig:limit_case_pil_tot}
\end{figure*}
The mechanical response of a nano-particle film resting on an infinitely extended substrate (Fig.~\ref{fig:geometry_scheme}) is here analysed assuming negligible intrinsic acoustic losses. For the sake of the following discussion three layers are defined: the NP film layer ($q<z<h$), the interfacial layer ($0<z<q$) and the semi-infinite substrate layer ($z<0$). The problem is considered one-dimensional, as is the case for photoacoustic measurements on ultrathin films \cite{ogi2011picosecond,peli2016mechanical,grossmann2017characterization}. The only non-zero component of the displacement field, $u_z^\#(z,t)$, satisfies the classic wave equation:
\begin{equation}
\dfrac{\partial^2 u_z^{\#} (z,t)}{\partial t^2} = v_z^{\#^2} \dfrac{\partial^2 u_z^{\#} (z,t)}{\partial z^2} \, ,
\label{eq:wave_equation}
\end{equation}
where $u_z^\# (z,t)$ is the displacement component in the $z$ direction, the hash refers to each layer, and $v_z^{\#}$ is the velocity of the P-wave travelling in such materials. The solution of Eq.(\ref{eq:wave_equation}) can be written as
\begin{equation}
u_z^{\#} (z,t) = U^{\#} (z)  T^{\#} (t) \, ,
\label{eq:variable_decomposition}
\end{equation}
with
\begin{align}
U^{\#} (z) &= u_k^{\#} e^{i k^{\#} z} + u_{-k}^{\#} e^{- i k^{\#} z} \, ,
\notag
\\
T^{\#} (t) &= u_\omega^{\#} e^{- i \omega t} \, ,
\label{eq:variable_decomposition_2}
\end{align}
where $i$, $\omega$, and $k^{\#}$ are the imaginary unit, the frequency and the wave vector, respectively.
Substituting Eq.(\ref{eq:variable_decomposition}) and (\ref{eq:variable_decomposition_2}) into Eq.(\ref{eq:wave_equation}) yields the dispersion relation $\omega^2 = v_z^{\#^2} k^{\#^2}$. The first and the second terms of $U^{\#} (z)$ are the regressive and the progressive components of the wave, respectively. The regressive component of the wave in the substrate is neglected since this layer is considered as infinitely extended in the $z$ direction, a fact accounting for the radiative attenuation of the film's breathing mode towards the substrate.

When dealing with granular solids, like the aforementioned nano-particle film (Fig.~\ref{fig:geometry_scheme},left), imposing a ``perfect adhesion'' condition (\textit{pa}) at the film-substrate interface results in a faulty evaluation of their mechanical behaviour.
Perfect adhesion implies perfect geometrical matching and continuity of stress and displacement.
This fault is particularly relevant when addressing the oscillation's damping time, not as much for the oscillation frequency \cite{peli2016mechanical}.
This can be traced back to the fact that granularity makes the perfect contact condition unlikely to be achieved, whereas a `patched interface' would be more appropriate.

To overcome this issue, the pillar model is introduced (Fig.~\ref{fig:geometry_scheme},centre). The pillar model partitions the nanogranular film of thickness $h$ (Fig.~\ref{fig:geometry_scheme}, left) in three layers (Fig.~\ref{fig:geometry_scheme}, centre). The actual NP film layer, $q<z<h$, is accounted for introducing an \textit{effective} homogeneous and isotropic thin film layer extending in the same range. The real NP film morphology is granular rather than homogeneous, nevertheless simulating the real NP film with an homogeneous one allows defining an \textit{effective} density $\rho^{NP}$ and an \textit{effective} stiffness tensor $C^{NP}$. These constants may be retrieved either from experiments \cite{peli2016mechanical} or theory \cite{benetti2017bottom,benetti2018photoacoustic}. The key element in the model is the introduction of a layer of pillars (dashed orange layer in Fig.~\ref{fig:geometry_scheme}, centre), extending in the range $0<z<q$ and intended to mimic the mechanics in the  interfacial layer, i.e. at the interface between the actual film and the substrate (dashed orange layer in Fig.~\ref{fig:geometry_scheme}, left). The pillars density, $\rho^{bk}$, and Young modulus, $E^{bk}$, are taken as the ones of the real material of which the NPs are made of. The pillar mechanical properties hence differ from that of the effective NP thin film layer. The pillar layer adheres on a semi-infinite substrate, $z<0$.

The velocity, $v_z^{NP}$, of a P-wave in the NP film layer is proportional to the coefficient $C_{11}^{NP}$ since transversal contraction is prevented:
\begin{equation}
v_z^{NP} = \sqrt{\dfrac{C_{11}^{NP}}{\rho^{NP}}} \, ,
\label{eq:vlocityNP}
\end{equation}
while the velocity of a P-wave in the pillars is proportional to the Young modulus $E^{bk}$ since they are free to expand transversely:
\begin{equation}
v_z^{NP} = \sqrt{\dfrac{E^{bk}}{\rho^{bk}}} \, ,
\label{eq:vlocityPil}
\end{equation}

For the pillar model, the boundary conditions are the following:
\begin{enumerate}
\item free standing at the top of the effective NP-layer ($z=h$):
\begin{equation}
C_{11}^{NP} \dfrac{\partial u_z^{NP} \left(h,t\right)}{\partial z} = 0 \, ,
\label{eq:BCP1}
\end{equation}
\item equilibrium at the interface between the effective NP-layer and the pillars layer ($z=q$):
\begin{equation}
F^{NP}\left(q,t\right) = F^{pil}\left(q,t\right) \, ,
\label{eq:BCP2}
\end{equation}
\item continuity of the displacement at the interface between the effective NP-layer and the pillars layer:
\begin{equation}
u_z^{NP}\left(q,t\right) = u_z^{pil}\left(q,t\right) \, ,
\label{eq:BCP3}
\end{equation}
\item force equilibrium at the interface between the pillars layer and the sapphire substrate ($z=0$):
\begin{equation}
F^{pil}\left(0,t\right) = F^{sub}\left(0,t\right) \, ,
\label{eq:BCP4}
\end{equation}
\item continuity of the displacement at the interface between the pillars layer and the sapphire substrate:
\begin{equation}
u_z^{pil}\left(0,t\right) = u_z^{sub}\left(0,t\right) \, ,
\label{eq:BCP5}
\end{equation}
\end{enumerate}
It is pinpointed that the continuity of the stresses at the interfaces between the pillars and the two continuous layers is replaced with the balance of their resultant forces, $F$, as can been appreciated in Eq.(\ref{eq:BCP2}) and Eq.(\ref{eq:BCP4}). This is a key point of the model.\\
Eq.(\ref{eq:BCP2}) and Eq.(\ref{eq:BCP4}) reduce to
\begin{align}
C_{11}^{NP} \dfrac{\partial u_z^{NP} \left(q,t\right)}{\partial z} &= \alpha E^{bk} \dfrac{\partial u_z^{pil} \left(q,t\right)}{\partial z} \, ,
\notag
\\
\alpha E^{bk} \dfrac{\partial u_z^{pil} \left(0,t\right)}{\partial z} &= C_{11}^{sub} \dfrac{\partial u_z^{sub} \left(0,t\right)}{\partial z} \, ,
\label{eq:BCP_forces}
\end{align}
respectively, where $\alpha$ is the contact ratio between the areas of the two homogeneous layers (substrate and effective NP film) and that of the pillars  (see Fig.~\ref{fig:geometry_scheme}, right), $C_{11}^{sub}$ and $C_{11}^{NP}$ the substrate's and the effective NPs film relevant stiffness tensor's component, respectively.
The model is therefore reduced to 1D. 
It is noteworthy that, despite the fact that the pillars in Fig.~\ref{fig:geometry_scheme} are represented with a circular cross-section, the definition of the parameter $\alpha$ and the structure of the model do not change if the shape of such cross-section is chosen to be different, for instance square-shaped rather than circular. Further on, the analytical model does not depend on the position of the pillar with respect to the unit cell, this despite the fact that the pillars in Fig.~\ref{fig:geometry_scheme} are shown at its center. These two aspects are crucial for a model intended to correctly rationalize a disordered interface, were the number of possible NPs dispositions at the interface, i.e. number of micro-states or configuration in statistical mechanics terms, is infinite. In photo-acoustic experiments for instance, where both the excitation and probing laser beams are much wider than the NP's dimensions, a huge number of possible unit cell's configurations are probed all-together within a single measurement. The acoustic problem is therefore not affected by the specific global interface configuration
, hence for the pillar model to correctly capture the physics it must not depend on the specific pillar cross-sectional geometry or position within the unit cell.
\\
Enforcing the boundary conditions Eqs.(\ref{eq:BCP1})-(\ref{eq:BCP2})-(\ref{eq:BCP3})-(\ref{eq:BCP4})-(\ref{eq:BCP5}) in Eqs.(\ref{eq:variable_decomposition}) yields the following equation in the complex-valued unknown $\omega (q,\alpha)$:
\begin{widetext}
\begin{equation}
Z^{NP} -
\dfrac{\alpha  E^{bk} \cot \left(\dfrac{(h-q)\omega
}{v_{z}^{NP}}\right) \left[v_{z}^{pil}
Z^{sub} \cos \left(\dfrac{q~\omega
}{v_{z}^{pil}}\right)-i \alpha  E^{bk} \sin
\left(\dfrac{q~\omega
}{v_{z}^{pil}}\right)\right]}{v_{z}^{pil}
\left[v_{z}^{pil} Z^{sub} \sin \left(\dfrac{q~
\omega }{v_{z}^{pil}}\right)+i \alpha  E^{bk}
\cos \left(\dfrac{q~\omega
}{v_{z}^{pil}}\right)\right]}
= 0 \, ,
\label{eq:impli_solu_pil}
\end{equation}
\end{widetext}
where $Z^{NP}$, and $Z^{sub}$ are the acoustic impedances of the effective NP-layer and the substrate, respectively. Actually, Eq.(\ref{eq:impli_solu_pil}) may be solved numerically and yields, for each fixed set of parameters $(q,\alpha)$, infinitely many solutions $\omega = \omega_n (q,\alpha)$ with $n$=$\{$0,1,2,...$\}$ the index numbering the mode.

The total thickness $h$ of the actual real film assigned, the free parameter in Eq.(\ref{eq:impli_solu_pil}) are the height of the pillars, $q$, and the contact ratio between the two homogeneous layers and the pillars, $\alpha$. The relations linking the period of vibration, $T_n(q,\alpha)$, and the wave decay time, $\tau_n(q,\alpha)$, to the n-mode complex-valued angular frequency are:
\begin{align}
T_n(q,\alpha) = \dfrac{2 \pi}{Re(\omega_n(q,\alpha))} \, ,
\notag
\\
\tau_n(q,\alpha) = \dfrac{1}{Im(\omega_n(q,\alpha))} \, .
\label{eq:relation_period_decay}
\end{align}

The intuitive idea underlying the pillar model stands in the possibility to reduce the full 3D acoustic scattering problem, involving a disordered interface, to a more amenable 1D one. This approximation is meaningful provided the detailed distribution of stresses across the interface does not affect the solution in terms of quasi-mode period and lifetime. As a matter of fact, the 1D model retains information on the integral of the stresses exchanged across the interfaces rather than their detailed distribution. This key point finds its microscopic justification on the fact that the acoustic problem is not affected by the specific interface configuration, as earlier addressed.

The pillar model is more evolved with respect to spring-based interface models, which are commonly exploited to mimic imperfect interfaces, see for instance the seminal work of Ref. \cite{bigoni2002statics}. In the present case, the pillar has rigidity $\alpha E^{bk}L^{2}/q$, which, contrary to the spring rigidity, arises from the specific interface geometrical and physical characteristics. Furthermore, the pillars correctly account for inertia, the mass being distributed as opposed to concentrated, as is the case for mass-spring interface models and alike.
\subsection{The Pillar model: Case Study}
\begin{figure*}
  \centering
  \begin{minipage}[b]{\columnwidth}
    \includegraphics[width=\columnwidth]{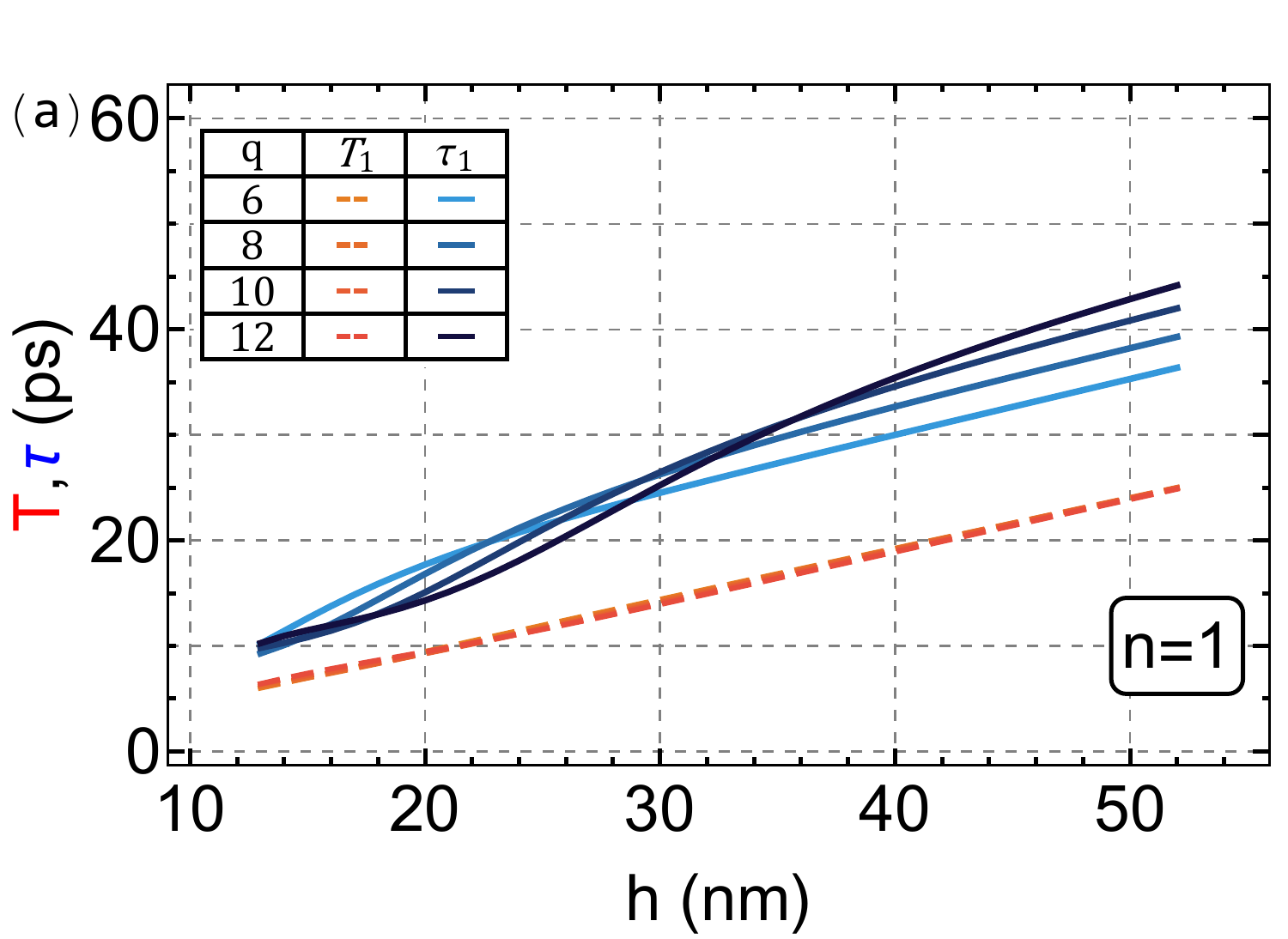}
  \end{minipage}
  \hfill
  \begin{minipage}[b]{\columnwidth}
    \includegraphics[width=\columnwidth]{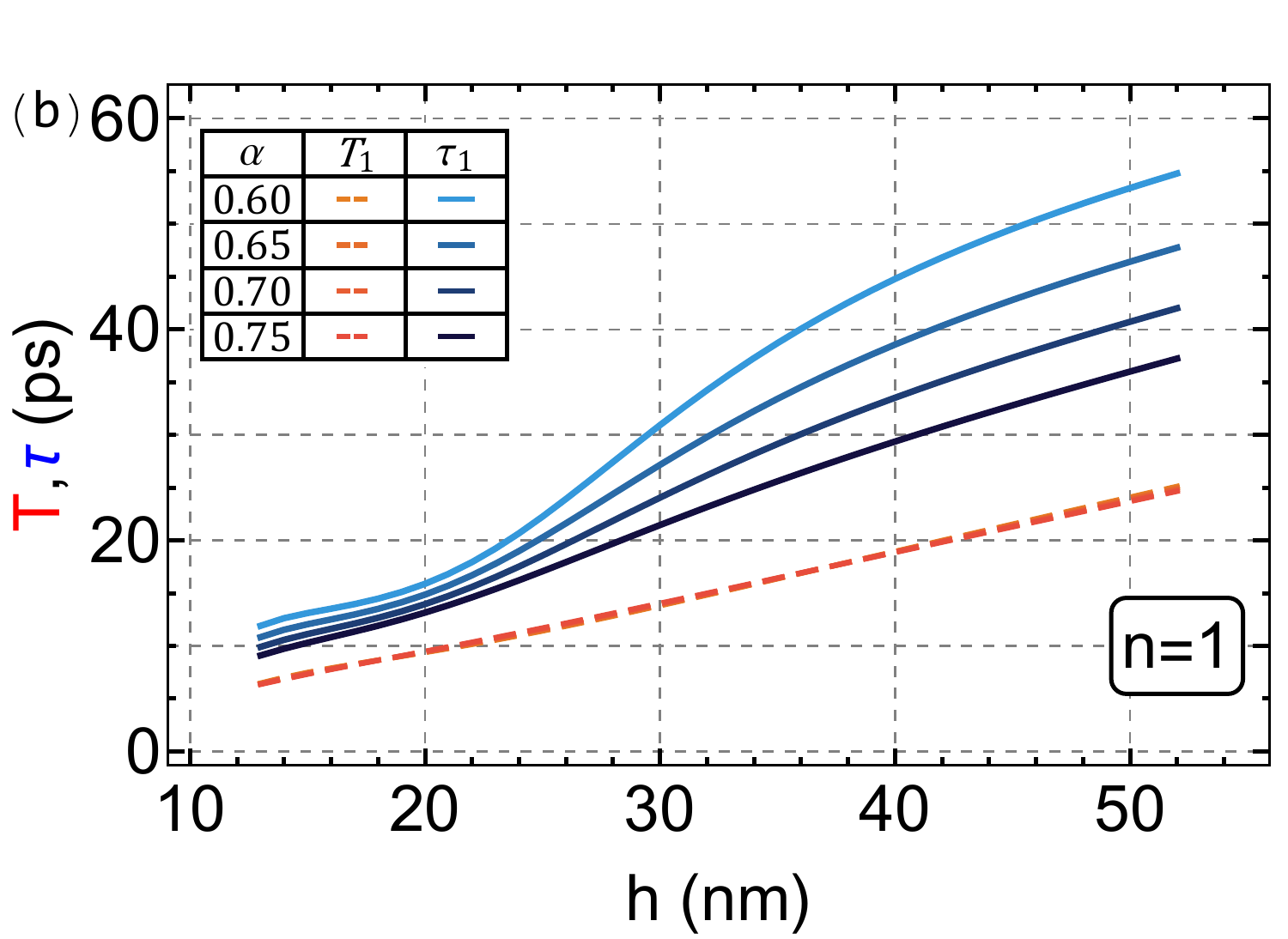}
  \end{minipage}
  \caption{$T_1(h;q,\alpha)$ and $\tau_1(h;q,\alpha)$ versus $h$ for $n=1$ for the Ag nanogranular film:
    (a) fixed $\alpha =  0.68$ while varying $q$ (expressed in nm);
    (b) fixed $q = 12$ nm for a limited span of $\alpha$ values centred around the best fitting value $\alpha =  0.68$;
    The plots of $T_1(h;q,\alpha)$ (dashed line, orange colour range) graphically overlap, not so for $\tau_1(h;q,\alpha)$ (continuous line, blue colour range).}
  \label{fig:parametric_study_pil_tot}
  \end{figure*}
The pillar model is here exemplified for the case of a real granular thin film \cite{peli2016mechanical} made of pure Ag NPs $\sim$ 6 nm in diameter, total film thickness $h$=50 nm, filling factor 0.8 and adhered on a sapphire substrate, (0001) $\alpha$-Al$_{2}$O$_{3}$ single crystal, of acoustic impedance $Z^{sub}$. Acoustic damping was shown to be due to extrinsic losses, a condition that must be met in order for Eq.(\ref{eq:impli_solu_pil}) to be applicable. The NP film is well mimicked by an homogeneous effective film of known mechanical properties: $v_{z}^{NP}$, $\rho^{NP}$ and $Z^{NP}$. The concept of NP film is meaningful beyond the first two deposited layers of NPs, leading to an interface layer of $\sim$12 nm, as detailed in Ref \cite{benetti2017bottom}. A value of $q$=12 nm is therefore assumed for the pillars, which are made of pure Ag of density $\rho^{bk}$, Young modulus $E^{bk}$ and sustain P-waves of sound velocity $v_{z}^{bk}$. The pillar layer filling factor $\alpha$ is here left as the sole free parameter, Eq.(\ref{eq:impli_solu_pil}) thus linking the complex-valued unknown $\omega$ to $\alpha$. The values of the relevant mechanical properties for this system are reported in table \ref{tab:para_peli}.

The oscillation period $T_{n}$ and lifetime $\tau_{n}$ for the first two modes of the pillar model, $n$=$\{0,1\}$, are reported versus $\alpha$ as full lines in Fig.~\ref{fig:limit_case_pil_tot} panel (a) and (b), respectively. For $\alpha=0.8$ the density of the pillars layer matches the density of the NP-layer, the latter being 0.8 that of bulk Ag. Densification of the interface layer with respect to the NP film's bulk was ruled out for the present scenario \cite{benetti2017bottom}, the maximum value of $\alpha$ is hence here constrained to 0.8. A comment is here due. For the case of cylindrical pillars, a value of $\alpha$$>$$\pi/4\approx0.78$ implies compenetration of neighbouring pillars. This fact does not constitute a problem though, since, as previously discussed, the model is independent on the pillar's cross-section geometry. For instance, for a pillar of square cross-section, compenetration is prevented for any value of $\alpha$$<$1.
For the pillar model, the period and lifetime of a given mode $n$=$\{0,1,2,...\}$ (with $n$=0 meaning $n\to0$) are correctly bounded between those of a ``free standing" ($\textit{fs}$) NP film of thickness $h-q$:
\begin{equation}
\begin{cases}
T_{n,fs}(\alpha) = \dfrac{2 (h-q)}{v_{z}^{NP}}\dfrac{1}{n} \, ,
\qquad\,
n=$\{0,1,2,...\}$
\\
\tau_{n,fs}(\alpha) = \infty $ \, ,
\qquad\qquad\quad\,\!\!
$ \forall n
\end{cases}
\quad
\label{eq:free_T_tau}
\end{equation}
and those of the ``perfect adhesion" ($\textit{pa}$) model:
\begin{equation}
\quad
\begin{cases}
T_{n,pa}(\alpha) = \dfrac{4 h}{v_{z}^{NP}}\dfrac{1}{\left(1+2n\right)} \, ,
\\
\tau_{n,pa}(\alpha) = \dfrac{2 h}{v_{z}^{NP}}\left|ln\left(\left|\dfrac{Z^{sub}-Z^{NP}}{Z^{sub}+Z^{NP}}\right|\right)\right|^{-1} \, .
\end{cases}
\label{eq:perf_T_tau}
\end{equation}
Attention is drawn on the fact that actually $\tau_{n,pa}(\alpha)$ is mode independent.
Indeed, as $\alpha$ approaches zero so does the pillars cross-sectional area and the pillar model converges to that of a free standing NP film of thickness $h-q$. On the contrary, as $\alpha$ approaches one, and assuming a square cross-section for the pillars, the situation converges to that of a perfectly adhering film (continuity of displacement and normal stress component at the interface) of thickness $h$.
Specifically, for $n$=0, $T_0$ diverges (meaning a film rigid shift) as $\alpha$ approaches zero, as expected for the $\textit{fs}$ film, and is 70 ps for $\alpha$=0.8, that is converging to the period of the fundamental mode, $T_{0,pa}$, for the $\textit{pa}$ case, see Fig.~\ref{fig:limit_case_pil_tot}(a). On the same footing, the mode lifetime $\tau_0$ diverges upon approaching the $\textit{fs}$ limit, whereas it approaches the lifetime of the $\textit{pa}$ film, $\tau_{pa}\sim$30 ps, for $\alpha$=0.8, regardless of the specific mode, see Fig.~\ref{fig:limit_case_pil_tot}(b). For $n$=1, $T_1$ evolves from $T_{1,fs}$=27 ps, for $\alpha$=0, to close to $T_{1,pa}$=23 ps, for $\alpha$=0.8.
The small gap between $T_{1}$ and $T_{1,pa}$ is due to the fact that wave propagation is governed by $E^{bk}$ in the pillars layer and by $C^{NP}_{11}$ in the NP film, see Fig.~\ref{fig:limit_case_pil_tot}(a).
As for the lifetime, $\tau_{1}$ qualitatively behaves as $\tau_{0}$ with respect to the $\textit{fs}$ and $\textit{pa}$ cases.
Interestingly, $\tau_{1}>\tau_{0}$ over the entire range of $\alpha$ values, see Fig.~\ref{fig:limit_case_pil_tot}(b).

The present discussion clearly demonstrates that the mode lifetime, rather than its oscillation period, is mostly sensitive to the interface morphology.
For instance, with reference to $n$=1, varying $\alpha$ so as to evolve from the $\textit{fs}$ to the $\textit{pa}$ film, the relative variation in the oscillation period is $\Delta T_{1}/T_{1}$$\sim$17$\%$ whereas the relative variation in lifetime $\Delta \tau_{1}/\tau_{1}$ is infinite.
This also explains why, in previous photo-acoustics experiments performed on granular thin films, the $\textit{pa}$ model was able to correctly address, within the error bar, the breathing mode oscillation period but failed in reproducing the lifetime \cite{peli2016mechanical}. Furthermore, it shows that the pillar model behaves correctly reproducing the $\textit{fs}$ and $\textit{pa}$ cases.

\subsection{The Pillar model: Parametric Study}
Typically, when undertaking an acoustic or photoacoustic investigation of the mechanical properties of ultra-thin films, one measures the breathing mode period and lifetime of a specific mode, $n$, over varying film's thicknesses, $h$.
The interface layer morphology, accounted for by the interface layer filling factor, $\alpha$, and its thickness, $q$, may therefore be retrieved from fitting of the experimental data exploiting the pillar model.
It is therefore important to undertake a parametric study to inspect how the parameters $\alpha$ and $q$ affect $T_{n}(h)$ and $\tau_{n}(h)$.
The calculations are here performed assuming the mechanical properties of the granular NP film addressed above.
For sake of exemplification, we here focus on mode $n=1$, which was the best characterised mode in previous experimental work.
$T_1(h;q,\alpha)$ and $\tau_1(h;q,\alpha)$ are reported versus the total thickness $h$ of the NP-layer for a fixed value of $\alpha = 0.68$ (the value that gives optimal fitting of the photoacoustic data) while varying the parameter $q$ across the set of values $\{6,8,10, 12\}$ nm (Fig.~\ref{fig:parametric_study_pil_tot}(a)) and, vice versa, fixing a value of $q=12$ nm (the value that gives optimal fitting of the photoacoustic data) while varying $\alpha$ across the set of values $\{0.60,0.65,0.70, 0.75\}$ (Fig.~\ref{fig:parametric_study_pil_tot}(b)).
This set of values has been chosen around the best fitting value $\alpha$=0.68, arising from fitting the experimental data pertaining to the sample here addressed, as detailed further on. \textit{Within this parameters range} and with reference to $\tau_1(h;q,\alpha)$, the two parameters act rather independently, $q$ and $\alpha$ governing the position of the inflection point (see Fig.~\ref{fig:parametric_study_pil_tot}(a)) and the tangent at the very same point (see Fig.~\ref{fig:parametric_study_pil_tot}(b)), respectively. Indeed, for a fixed $\alpha$, the flex moves towards higher $h$ values as $q$ increases, whereas, for a fixed $q$, as $\alpha$ approaches unity, the tangent's slope decreases, attaining an asymptotic value concomitantly with the curvature reaching zero. Far enough from the inflection point, $\tau_1(h;\alpha)$ is rather linear with $h$ (see Fig.~\ref{fig:parametric_study_pil_tot}(b)). As for the periods $T_1(h;q,\alpha)$, the differences are not appreciable throughout the presently explored range, see Fig.~\ref{fig:parametric_study_pil_tot} dashed orange blend lines.

\begin{figure}[h!]
	\centering
	\includegraphics[width=\columnwidth]{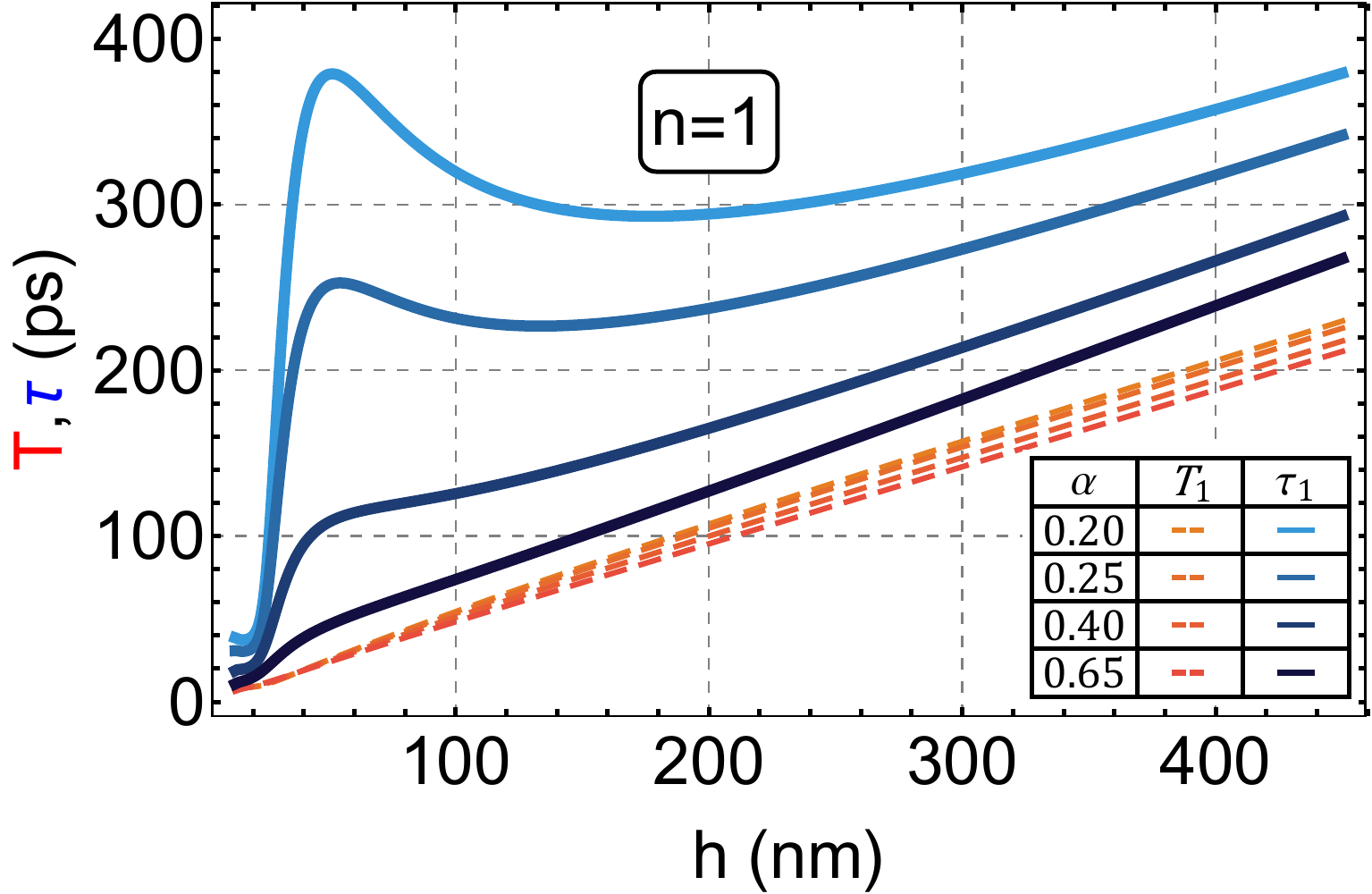}
	\caption{$T_1(h;q,\alpha)$ and $\tau_1(h;q,\alpha)$ versus an extended $h$ range for $n=1$ for the Ag nanogranular film: $q = 12$ nm and $\alpha$ values over an extended span.}
	\label{fig:parametric_study_pil_extended}
\end{figure}

Solutions obtained over a wider $\alpha$ and $h$ span are reported in Fig.~\ref{fig:parametric_study_pil_extended} for the same value of $q$=12 nm. Two features clearly arise. First, when extending the analysis to include also smaller $\alpha$, i.e slender pillars, a resonance in $\tau_1(h;q,\alpha)$ clearly emerges, and grows more pronounced as the pillar gets slender, see the decay times curve for $\alpha$ values of 0.4, 0.25, 0.2. 
This fact may be intuitively rationalized considering that, as the pillar gets slender, the situation approaches that of a free standing film. Formally, the pillar stiffness decrees proportionally to its cross section (that scales with $\alpha$), resulting in a monotonous reduction of the mechanical wave propagation speed, ultimately extending the quasi-mode life time.

These resonances stand out also in the mode's $Q$ factor, a feature recently observed also in the context of a single nanodisk adhered on a substrate \cite{medeghini2018controlling}. Secondly, for large enough values of $h$, that is once the pillar length becomes negligible with respect the total thickness of the nanoparticle film, $\tau_1(h;q,\alpha)$ scales rather linearly with $h$. In this $h$ range also the minute differences in the periods $T_1(h;q,\alpha)$ for different $\alpha$ values can be appreciated, see Fig.~\ref{fig:parametric_study_pil_extended} orange-blend curves.
\subsection{Pillar model benchmarking: fitting photoacoustic data}
The pillar model is now deployed to fit photoacoustic data acquired on nanogranular films of different thicknesses \cite{peli2016mechanical}.
The samples are the same as the one addressed in the case study. These samples constitute an ideal system for benchmarking purposes.
The peculiarities of the deposition method \cite{wegner2006cluster} allow to obtain solvent free and ultra-pure nanoporous films, avoiding the synthesis-related complicacies involved in other methods.
Furthermore, these films have been fully characterised in terms of compositional, structural, morphological and mechanical properties.
On a general basis, the interface layer properties are the one which prove harder to access.
Whereas the NP film layer filling factor may be retrieved employing a variety of techniques, such as X-ray reflectivity \cite{peli2016mechanical}, environmental ellipsometric porosimetry \cite{bisio2009optical} and combining the Brunauer-Emmett-Teller method (BET) with Atomic Force Microscopy (AFM) \cite{borghi2019quantitative}, the interface layer filling factor $\alpha$ and thickness $q$ escape direct inspection.
Only recently, were the latter quantities operatively defined and estimated via a combined Transmission Electro Microscopy (TEM) and Molecular Dynamic (MD) investigation performed on the samples here addressed \cite{benetti2017bottom}. Specifically, the interface layer thickness is defined as the minimal film thickness beyond which the slice filling factors, calculated for thicker films, overlap, as addressed in all details in \cite{benetti2017bottom}.

\begin{figure}
\centering
\includegraphics[width=\columnwidth]{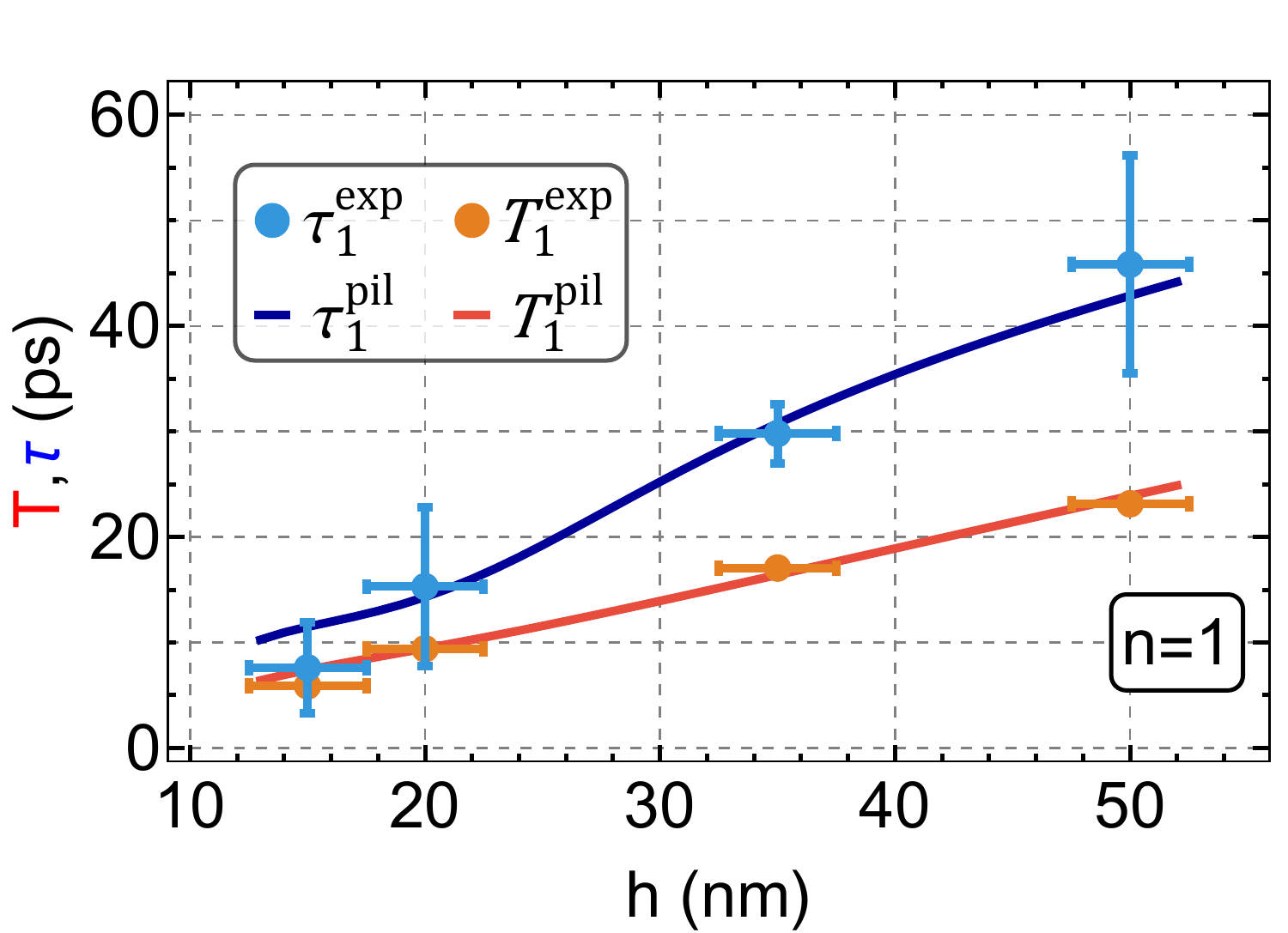}
\caption{Pillar model's best fit solution for mode $n$=1 for the Ag nanogranular film: $T_1(h;q,\alpha)$ (continuous orange line) and $\tau_1(h;q,\alpha)$ (continuous blue line) vs $h$ plotted for the best fit parameters $q=12$ nm and $\alpha=0.68$. The fitting is performed against the experimental data from \cite{peli2016mechanical}: $T_{1,exp}(h)$ (light orange dots) and $\tau_{1,exp}(h)$ (light blue dots). The error bars on the measured oscillation periods, although present, are too small to be appreciated.}
\label{fig:optimal_solution_pillar}
\end{figure}

The pillar model is benchmarked by letting $q$ and $\alpha$ as fitting parameters and maximising the likelihood between the $h$-dependent functions $T_1(h;q,\alpha)$ and $\tau_1(h;q,\alpha)$ and the experimental values, $T_{1,exp}(h)$ and $\tau_{1,exp}(h)$, reported  in \cite{peli2016mechanical}. Results are reported in Fig.~\ref{fig:optimal_solution_pillar} for the best fit values of $q=12$ nm and $\alpha=0.68$ (continuous lines) together with the experimental data (markers). Fitting eight data points with two free parameters may not be ideal, nevertheless, the best fit parameters are fully consistent with the values that have been retrieved by other means: $q$=12 nm and $\alpha$$\sim$0.7 for the interface layer \cite{benetti2017bottom}. This is to say that, in the fitting procedure, one could have taken $\alpha$ as the sole fitting parameter, or even fixed all the parameters from previous knowledge, still landing on the experimental data with the theoretical curves calculated adopting the pillar model. The value $\tau_{1}$($h$=15 nm) falls at the edge of the error bar of $\tau_{1,exp}$($h$=15 nm): for $h$=15 nm the effective NP layer is only 3 nm thick, approaching the limit where only an interface layer exists and the concept of a film becomes questionable. Summarising, the pillar model allows rationalizing the experimental data, the best fitting parameter being fully consistent with the values expected from previous knowledge.\\
\subsection{Pillar model benchmarking: 3D pillar model solved by FEM}
\begin{figure*}
\centering
\includegraphics[width=2\columnwidth]{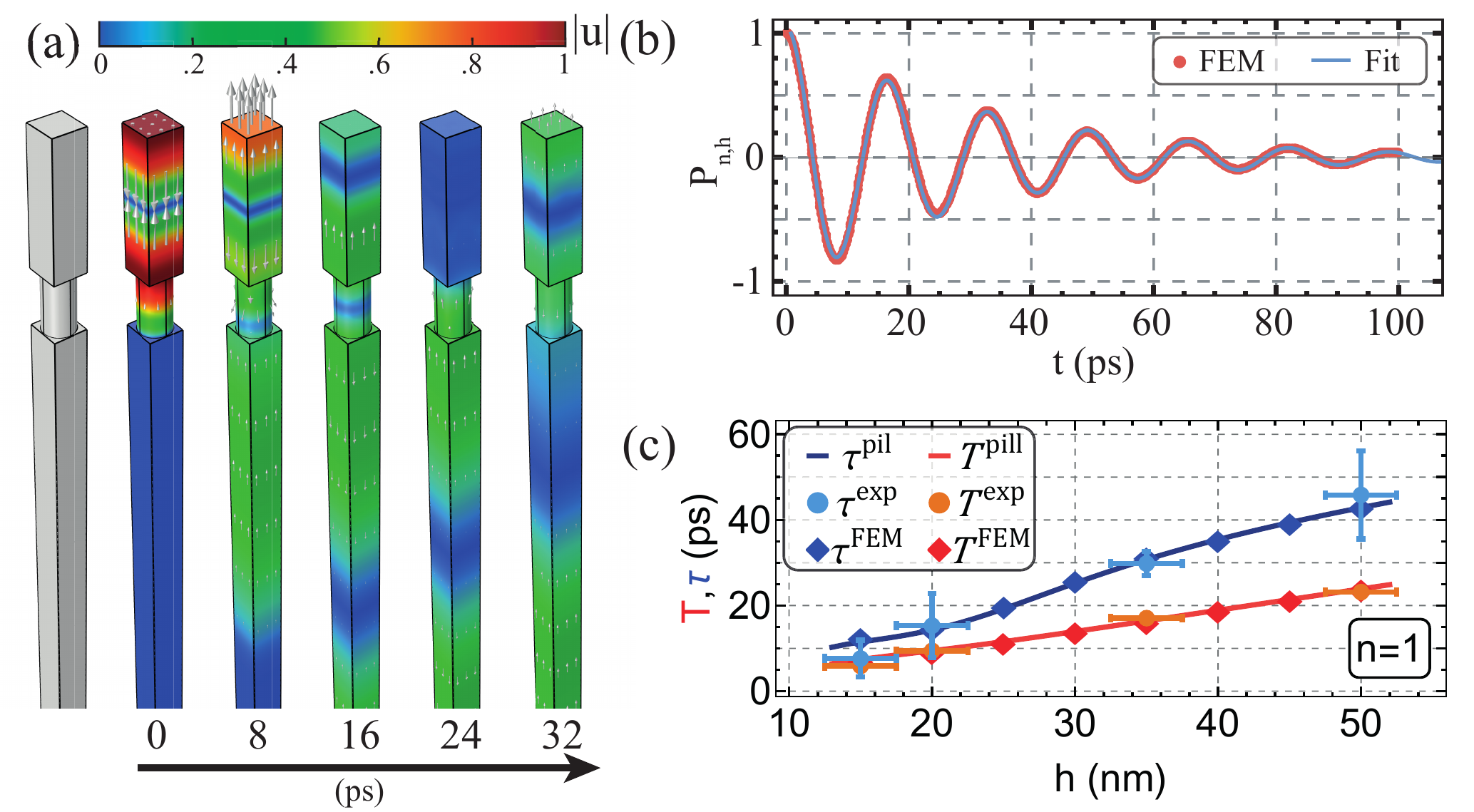}
\caption{Ag nanogranular film (see text).
(a) Simulation domain and displacement field $\textbf{u}_{n,h}(\textbf{r},t)$ (arrows) and modulus (colormap) at increasing times for $n$=1, $h$=40 nm and $q$= 12 nm. The displacement $\textbf{u}_{1,40}(\textbf{r},t=0)$ is constructed to match, for $z\ge0$, the film's eigenmode $n$=1.
(b) Normalized projection coefficient $P_{1,40}$ vs time for the case represented in panel (a) (full red dots); its fit with a damped oscillation of period $T$ and decay time $\tau$ (blue line).
(c) Periods and decay times vs film thickness: FEM simulations (diamonds), pillar model (solid lines) and experimental data from \cite{peli2016mechanical}(dots).
}
\label{fig:fem_simu}
\end{figure*}
We now compare the analytical 1D pillar model, addressed so far as the pillar model for brevity, against FEM simulations performed on the 3D pillar model.
The scope is twofold.
A first quest is whether the reduction from a full 3D pillar model (see Fig.~\ref{fig:geometry_scheme}, centre), where acoustic wave scattering is accounted for, to the 1D pillar model expressed by Eq.(\ref{eq:impli_solu_pil}), which does not account for scattering, is justified for the case of low $n$ modes.
Furthermore, the 3D model accounts for the distribution of stresses across the interfaces whereas the pillar model retains information on the integral of the stresses only.
Comparing results obtained from the pillar model against those of 3D FEM simulations would enable confirming the soundness of these approximations.
Secondly, although the pillar model benchmarked remarkably well against existing experimental data, the quest stands as whether the model remains effective across a wider range of interface layer filling factors (while keeping the NP film layer mechanical properties unaltered), a situation for which we lack experimental data.
In this sense comparing against FEM simulations constitute a valid alternative.

We then proceed as follows.
As a validation step, we first implement FEM simulations on the 3D pillar model, mimicking the situations for which experimental data are available.
That is we excite a specific film breathing mode, $n$= 1, and, subsequently, simulate its temporal evolution throughout the sample, now comprising the substrate as well, thus accessing the quasi-eigenmode oscillation period, lifetime and quality factor.
As a matter of fact, once the substrate is accounted for, the film breathing mode becomes a quasi-mode radiating acoustic energy into the substrate.
The results will be benchmarked against both those of the pillar model and the experimental ones.
We then run similar simulations varying the pillar layer filling fraction, $\alpha$, for a fixed film thickness, $h$=50 nm, and compare the results against the values obtained from the pillar model.

To this end we first consider the 3D pillar model (see Fig.~\ref{fig:geometry_scheme}, right) mimicking the samples on which experiments were performed and for which $q$=12 nm and $\alpha$=0.68 were obtained.

\noindent
{\textit{Geometry.}} The 3D unit cell geometry, reported in Fig.~\ref{fig:fem_simu}(a) and in right panel of Fig.~\ref{fig:geometry_scheme},
is composed of three domains and has base dimensions $L\times L$.
Domain \textit{`sub'} (-5 $\mu$m$<z<$0) consists in a 5 $\mu$m-thick sapphire substrate .
This value has been chosen long enough so as to avoid any wave front reflection from the bottom of sapphire within the time span of the simulated dynamics.
For the sake of visualization only a small part of it is shown.
Domain \textit{`pil'} (0$<z<$q) consists in a pure Ag cylindrical pillar of height $q$ and radius $r_{pil}=L \sqrt{\alpha/\pi}$.
We take $r_{pil}$=3.2 nm, consistent with the NPs radius composing the experimentally investigated films, thus resulting in $L$=7 nm.
Domain \textit{`NP'} (q$<z<$h) consists in the \textit{effective} NP layer of thickness $h-q$.\\
\begin{figure*}
	\centering
	\hspace{-0.5cm}
	\begin{minipage}[b]{0.68\columnwidth}
		\includegraphics[height=3.5cm]{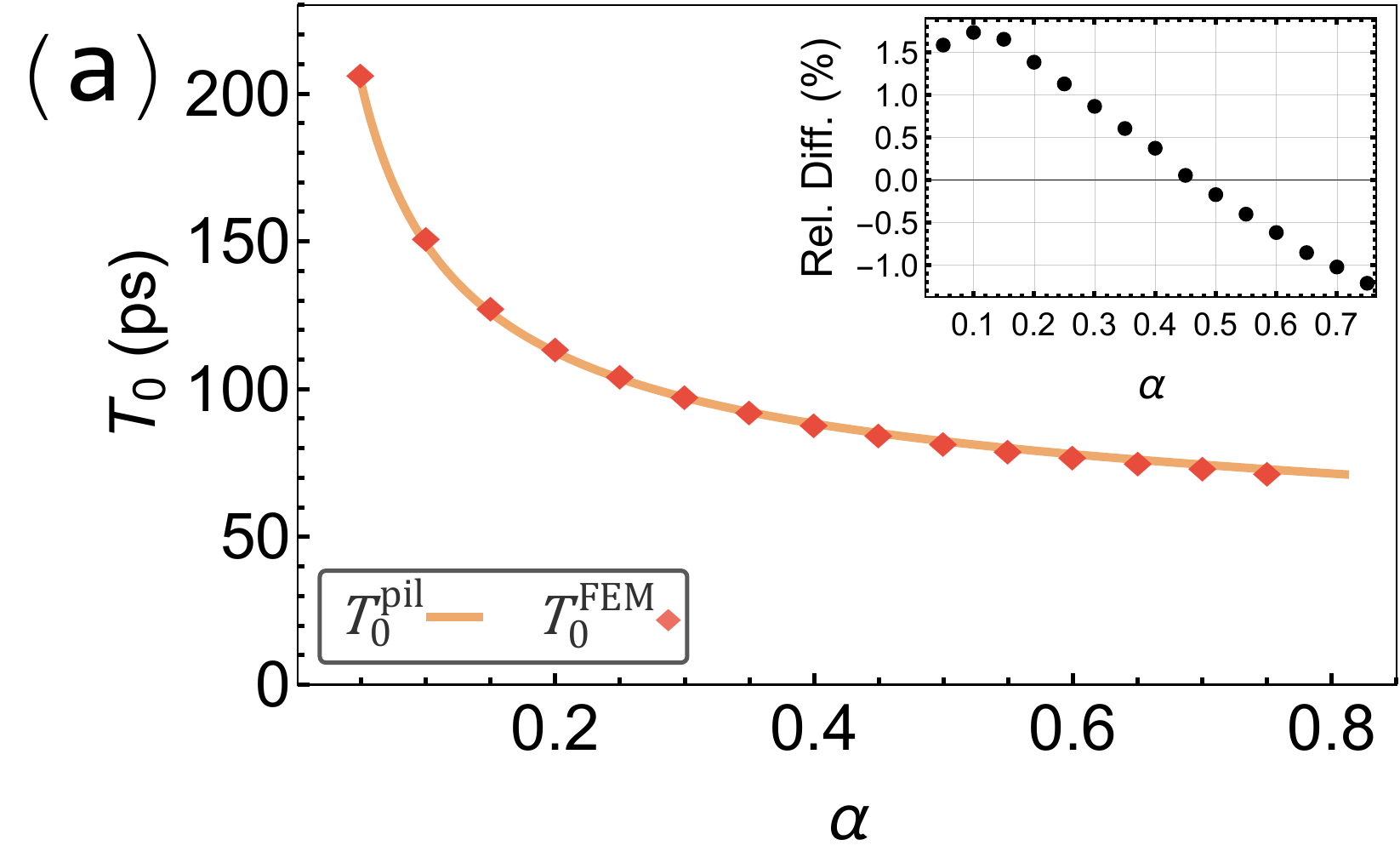}
	\end{minipage}
	\hfill
	\begin{minipage}[b]{0.68\columnwidth}
		\includegraphics[height=3.5cm]{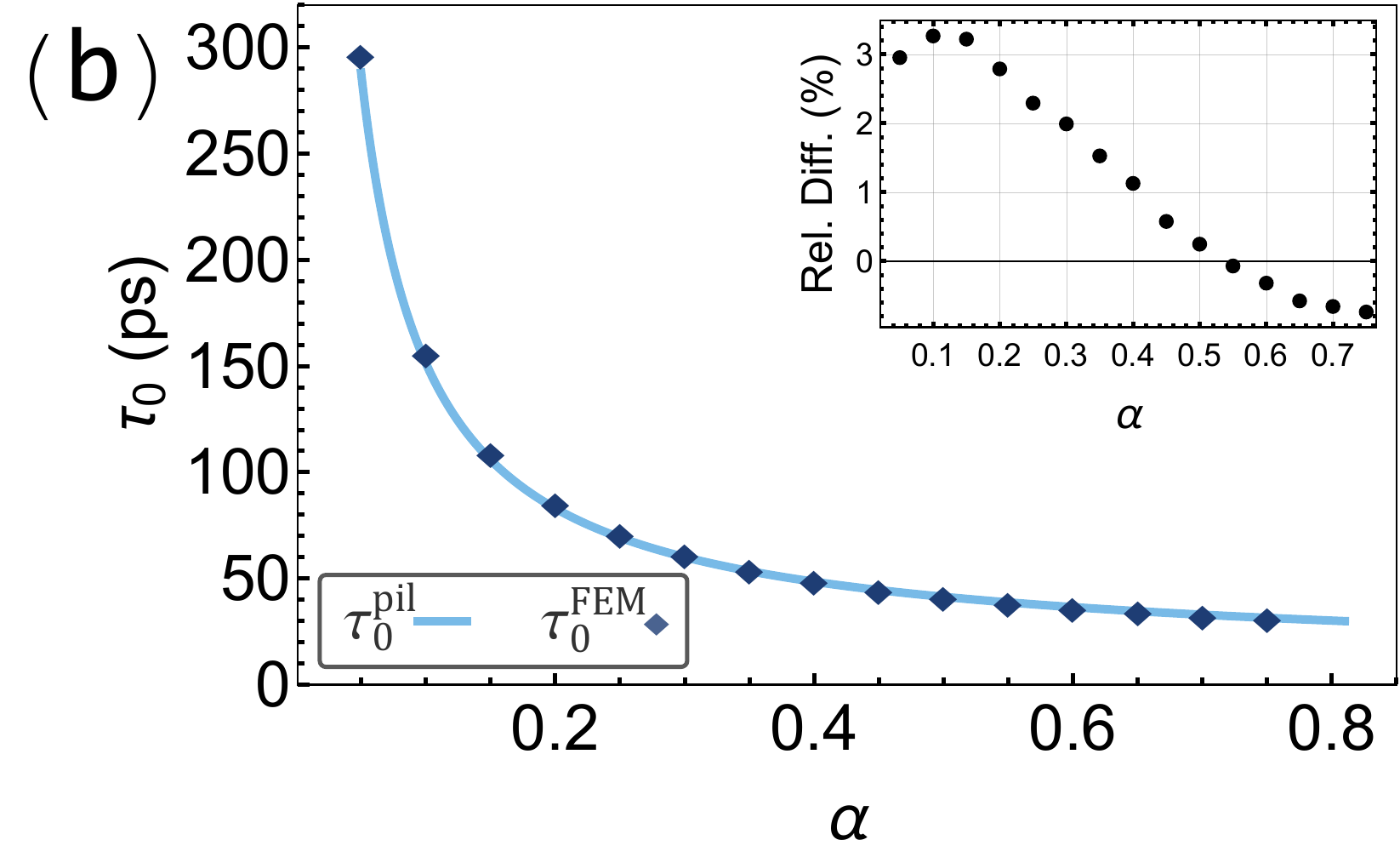}
	\end{minipage}
	\hfill
	\begin{minipage}[b]{0.68\columnwidth}
		\includegraphics[height=3.5cm]{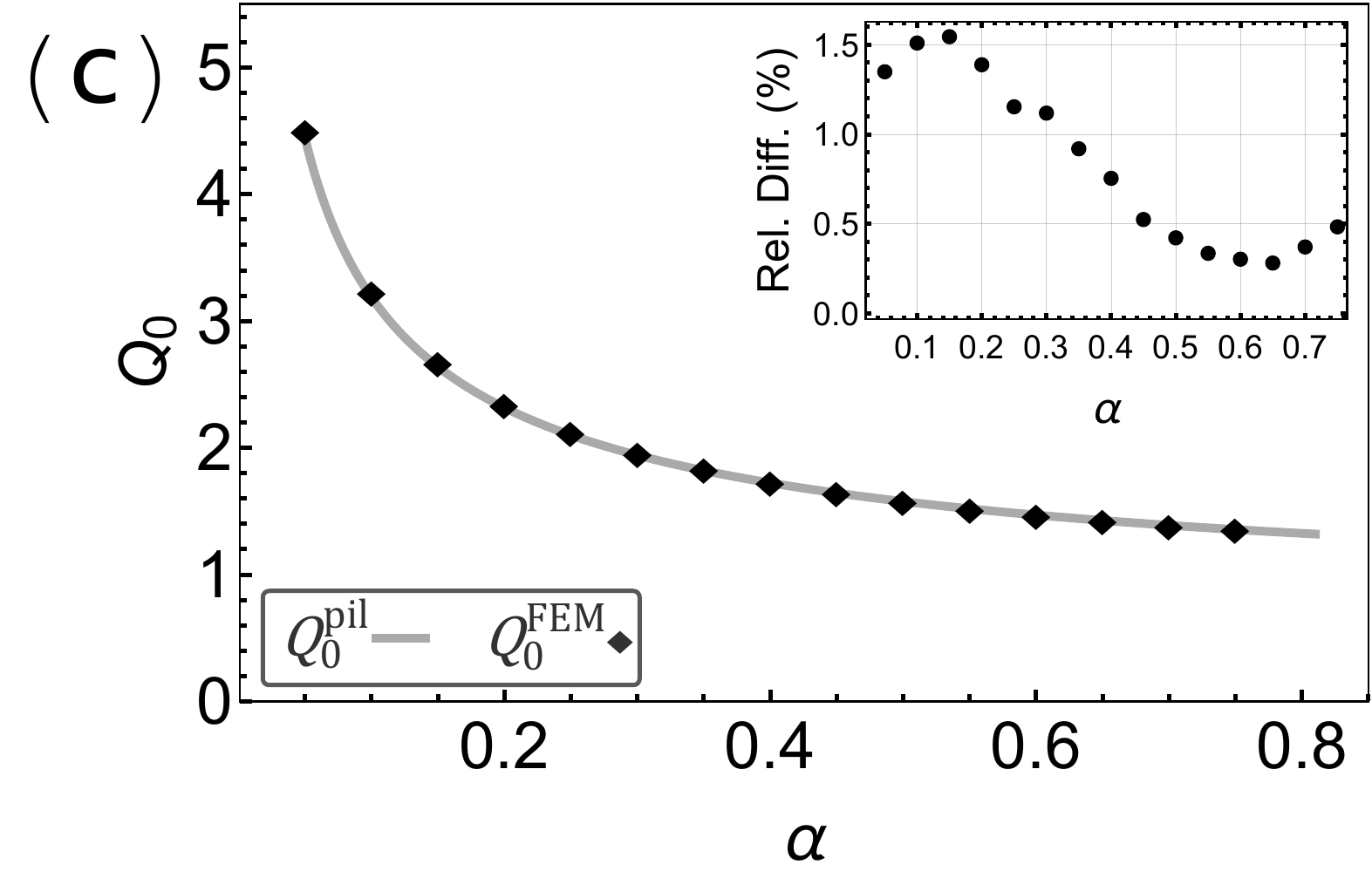}
	\end{minipage}
	\\
	\vspace{0.5cm}
	\hspace{-0.4cm}
	\begin{minipage}[b]{0.68\columnwidth}
		\includegraphics[height=3.5cm]{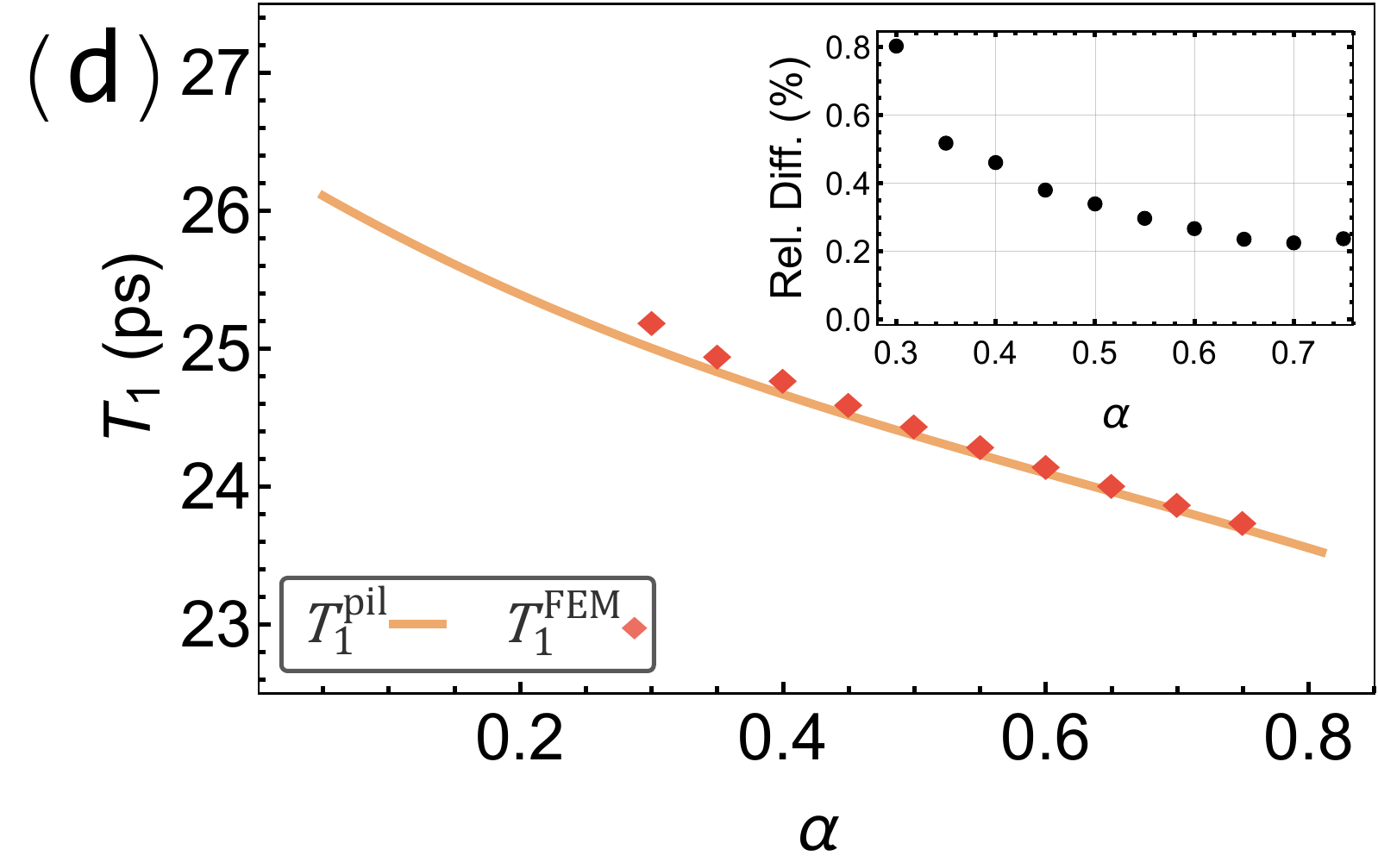}
	\end{minipage}
	\hfill
	\begin{minipage}[b]{0.68\columnwidth}
		\includegraphics[height=3.5cm]{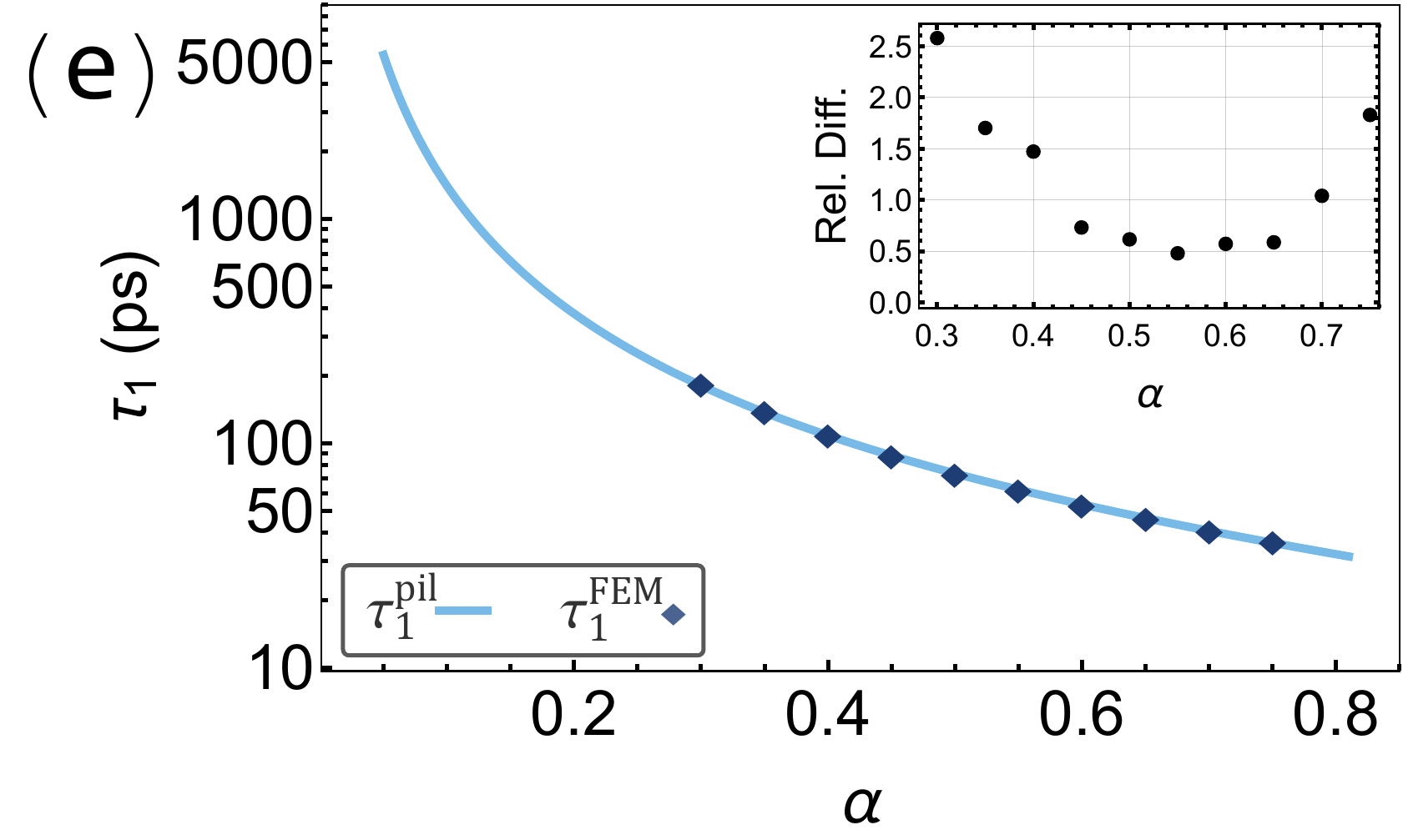}
	\end{minipage}
	\hfill
	\begin{minipage}[b]{0.68\columnwidth}
		\includegraphics[height=3.5cm]{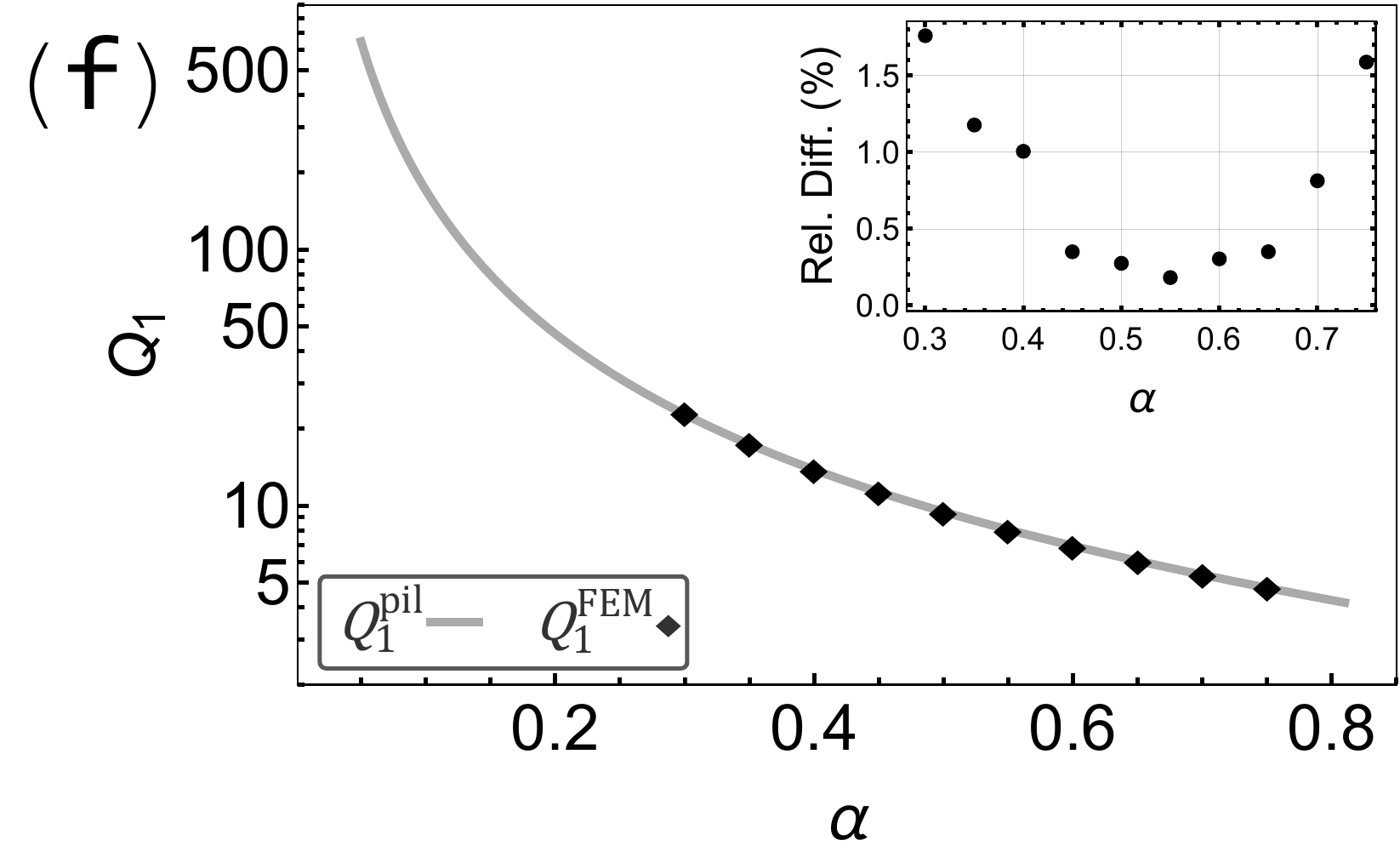}
	\end{minipage}
	\caption{Ag nanogranular film.
		Comparison between the oscillation period, decay time and quality factor obtained from FEM solution of the 3D pilar model (filled diamonds) and the pillar model (full lines) vs the pillar layer filling factor $\alpha$, for a film's thickness $h$=50 nm. (a) period, (b) decay time and (c) quality factor for $n$=0. (d) period, (e) decay time (log-lin scale) and (f) quality factor (log-lin scale) for $n$=1. The insets show the relative difference between the calculated quantities in the two models.
	}
	\label{fig:per_deca_quality_n_1}
\end{figure*}
{\textit{Materials properties.}} As for the domains mechanical properties, the densities and elastic constants for Sapphire and polycrystalline Ag are taken for the substrate and for the pillar, respectively, whereas the \textit{effective} NP layer is attributed the density $\rho^{NP}$ and the elastic tensor components $c_{11}$=6.96$\times$10$^{10}$ GPa, taken from \cite{peli2016mechanical}, and $c_{44}$=1,86 $\times$10$^{10}$ GPa, calculated from Budiansky homogenization formulas \cite{budiansky1965elastic} for a volumetric filling factor of 0.8.
The $c_{44}$ value is not actually of any relevance, since, given the problem's symmetry to be discussed shortly, the solution is independent on the choice of $c_{44}$, a fact that we numerically tested.
The adopted values for the above-mentioned quantities are reported in Table \ref{tab:para_peli}.\\
{\textit{Boundary conditions.}}
A zero-displacement boundary condition is enforced at the \textit{`sub'} bottom surface.
The \textit{`NP'} top surface is taken stress-free.
At the portion of the bottom surface of \textit{`NP'} not in contact with the pillar ($z$=$q^{+}$ and $\sqrt{x^2+y^2}>r_{pil}$), a stress-free boundary condition is enforced together with the constraint that the z-component of the displacement ( $w$ ) must be spatially constant along the $x-y$ plane (Rigid connector). Actually the rigid connector condition does not affect the result but slightly improves the computation time.
The displacement field component normal to the lateral boundaries of \textit{`sub'} and \textit{`NP'} is fixed to zero due to the system periodicity and the experimental excitation symmetry.
The pillar's wall is constrained to move in the vertical (i.e. the direction normal to the substrate) and radial direction only, so as to impede pillar torsion.
These boundary conditions have been chosen so as to be consistent with the pillar model.
Furthermore, the boundary conditions in both models, together with the irrelevance of the choice of $C_{44}$ in the domain \textit{`NP'}, are consistent with the displacement and stress fields symmetry triggered by an excitation mechanics such as that of a laser pulse, of waist much greater than the overall film thickness $h$ , impinging at normal incidence on the film.\\
{\textit{Film's quasi-mode period, life time and Q-factor.}} We first calculate the set of eigenmodes $\{\tilde{\textbf{u}}_{n,h}\left(\textbf{r}\right)\}$ solutions of the acoustic eigenvalue problem for the domain $\textit{`pil'} \, \cup \textit{`NP'}$ of height $h$:
\begin{equation}
\nabla\cdot \left[\textbf{c}(\textbf{r})\textbf{:}\nabla \tilde{\textbf{u}}_{n,h}(\textbf{r})\right] = -\rho(\textbf{r})\omega_i^2 \tilde{\textbf{u}}_{n,h}(\textbf{r})\;,\
\label{Acoustic_equation3}
\end{equation}
with $\rho\left(\textbf{r}\right)$ and $\textbf{c}\left(\textbf{r}\right)$  the position-dependent mass density and elastic stiffness tensor, respectively, and
with zero displacement enforced at the boundary $z$=0.
The latter is a good approximation for an impulsive excitation of the film (for instance upon absorption of an ultrashort laser pulse) when $Z_{sub}>Z_{pil}$, as in the present case.
The first subscript, $n$, identifies the film's eigenmode, the second, $h$, the film's thickness expressed in nm.

We then define the initial displacement on the entire simulation domain $\textit{`sub'} \, \cup \textit{`pil'} \, \cup \textit{`NP'}$:
\begin{equation}
\textbf{u}_{n,h}(\textbf{r},t=0)=\\
\begin{cases}
A\tilde{\textbf{u}}_{n,h}(\textbf{r})\, ,$    \quad\quad   $ \forall z\ge0
\\
0 \, , \hspace{1.4 cm} \quad\quad \forall z<0
\end{cases}
\quad
\label{Initial_displacement}
\end{equation}
%
\begin{figure*}
	\centering
	\includegraphics[width=2\columnwidth]{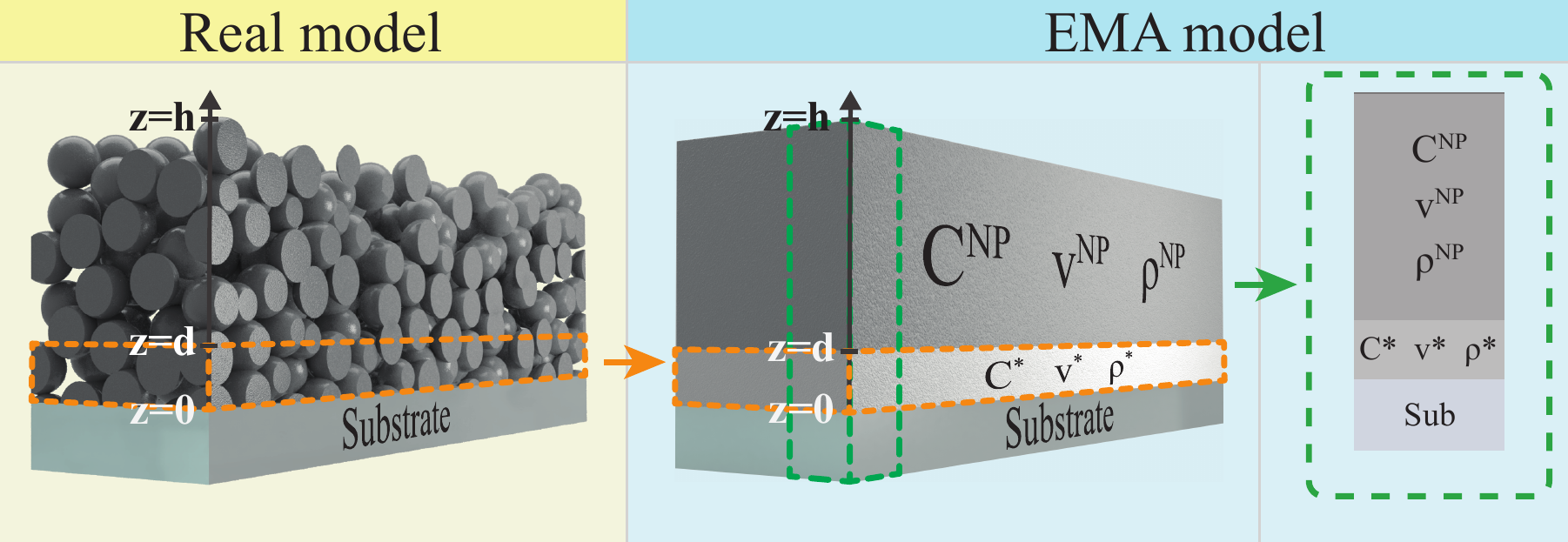}
	\caption{
		Left: 3D nanoparticles thin film of thickness $d$ adhered on a semi-infinite substrate. Centre: EMA model: \textit{effective} NP layer ($d<z<h$); \textit{effective} interface layer ($0<z<d$); semi-infinite substrate ($z<0$). The NP layer is the same one addressed in the pillar model. The interface layer has effective mechanical properties $C^{*}$, $v^{*}$ and $\rho^{*}$ (see text). The image is for illustrative purposes. Right: 1D sketch of the EMA model.
	}
	\label{fig:geometry_scheme_ema}
\end{figure*}
where the displacement amplitude $A$ will cancel out in the following analysis. 
We pinpoint that, $\textbf{u}_{n,h}(\textbf{r},t=0)$ \textit{is not} an eigenmode of the acoustic eigenvalue problem for the domain $\textit{`sub'} \, \cup \textit{`pil'} \, \cup \textit{`NP'}$, nevertheless, for $h\ge0$, it matches the eigenmode of domain $\textit{`pil'} \, \cup \textit{`NP'}$.
The initial velocity field is $\dot{\textbf{u}}_{n,h}(\textbf{r},t=0)=0$ .

Propagating the initial displacement on the entire unit cell via the Navier equation,
\begin{equation}
\nabla\cdot \left[\textbf{c}(\textbf{r})\textbf{:}\nabla \textbf{u}\right] = \rho\left(\textbf{r}\right)\ddot{\textbf{u}} \, ,
\label{Acoustic_equation_tdep}
\end{equation}
we obtain $\textbf{u}_{n,h}(\textbf{r},t)$.

For the sake of retrieving the film's quasi-eigenmode decay time we calculate the normalized projection coefficient between modes $\textbf{u}_{n,h}(\textbf{r},t=0)$ and $\textbf{u}_{n,h}(\textbf{r},t)$:
\begin{widetext}
\begin{equation}
P_{n,h} (t)=\frac{\langle \textbf{u}_{n,h}(t=0)|\textbf{u}_{n,h}(t) \rangle}{\langle \textbf{u}_{n,h}(t=0)|\textbf{u}_{n,h}(t=0) \rangle} =
\frac{\int_{V} \textbf{u}_{n,h}(\mathbf{r},t=0) \rho(\mathbf{r}) \textbf{u}_{n,h}(\mathbf{r},t) d\mathbf{r}}{\int_{V} \textbf{u}_{n,h}(\mathbf{r},t=0) \rho(\mathbf{r}) \textbf{u}_{n,h}(\mathbf{r},t=0) d\mathbf{r}} \, ,
\end{equation}
\end{widetext}
\noindent
where the integrals are actually calculated on the film's volume, $V_{film}$, since the initial displacement in the substrate is null by construction.
The introduction of the film density $\rho(\mathbf{r})$ is necessary to obtain a formally correct definition of the scalar product, the eigenvalue problem on the entire domain being of the Sturm-Liouville type.\\
For instance, for the case of a sample with $h$=40 nm and focusing on $n$=1, Fig.~\ref{fig:fem_simu} (a) shows the spatial profile of $\textbf{u}_{1,40}(\textbf{r},t=0)$ (arrows) and its modulus (colorbar) together with snapshots of its evolution $\textbf{u}_{1,40}(\textbf{r},t)$ taken for increasing times.
As time evolves, the film's quasi-eigenmode, $\textbf{u}_{1,40}(\textbf{r},t=0)$, fades away, displacement radiating into the substrate.
Fig.~\ref{fig:fem_simu} (b) reports the corresponding $P_{1,40}\left(t\right)$ (full red dots), measuring the overlap between the film's $n$=1 mode displacement profile at time $t$=0 and the actual displacement through out the sample at any given time $t$.
For the `gedanken' case, in which no acoustic radiation to the substrate occurs, the normalized projection coefficient would oscillate inbetween 1 and -1 without any damping, $\textbf{u}_{1,40}(\textbf{r},t)$ representing, for $z\ge$0, the film's quasi-eigenmode displacement at different times.
The normalized projection coefficient's maximum would thus be attained for $t=mT$ (the two displacements fields being in phase), it would be zero for $t=(2m+1)T/4$ (the two displacements fields being in quadrature) and be at its minimum for $t=(2m+1)T/2$ (the two displacements fields being in anti-phase) with $m\in \mathbb{N}_{0}$.
For the real case, in which acoustic radiation is active, the normalized projection coefficient's oscillation is exponentially damped, its period $T_{1,40}$ and decay time $\tau_{1,40}$ being retrieved fitting the numerical results, see Fig.~\ref{fig:fem_simu} (b), blue line.
Running simulations for varying $h$ we thus obtain $T_{n,h}$ and $\tau_{n,h}$, Fig.~\ref{fig:fem_simu} (c) reporting the case for $n$=1 (filled diamonds).
For the sake of comparison, we report on the same graph the data obtained from the analytic solution of the pillar model (full lines) together with the experimental values from ultrafast optoacoustic measurements \cite{peli2016mechanical} (filled circles).
The three sets of date are in good agreement, pointing to the fact that we correctly addressed the 3D pillar model via FEM and that, at least for $\alpha$=0.68, the approximations entailed in the pillar model are sound.
\begin{figure*}
	\centering
	\begin{minipage}[b]{1.03\columnwidth}
		\includegraphics[width=\columnwidth]{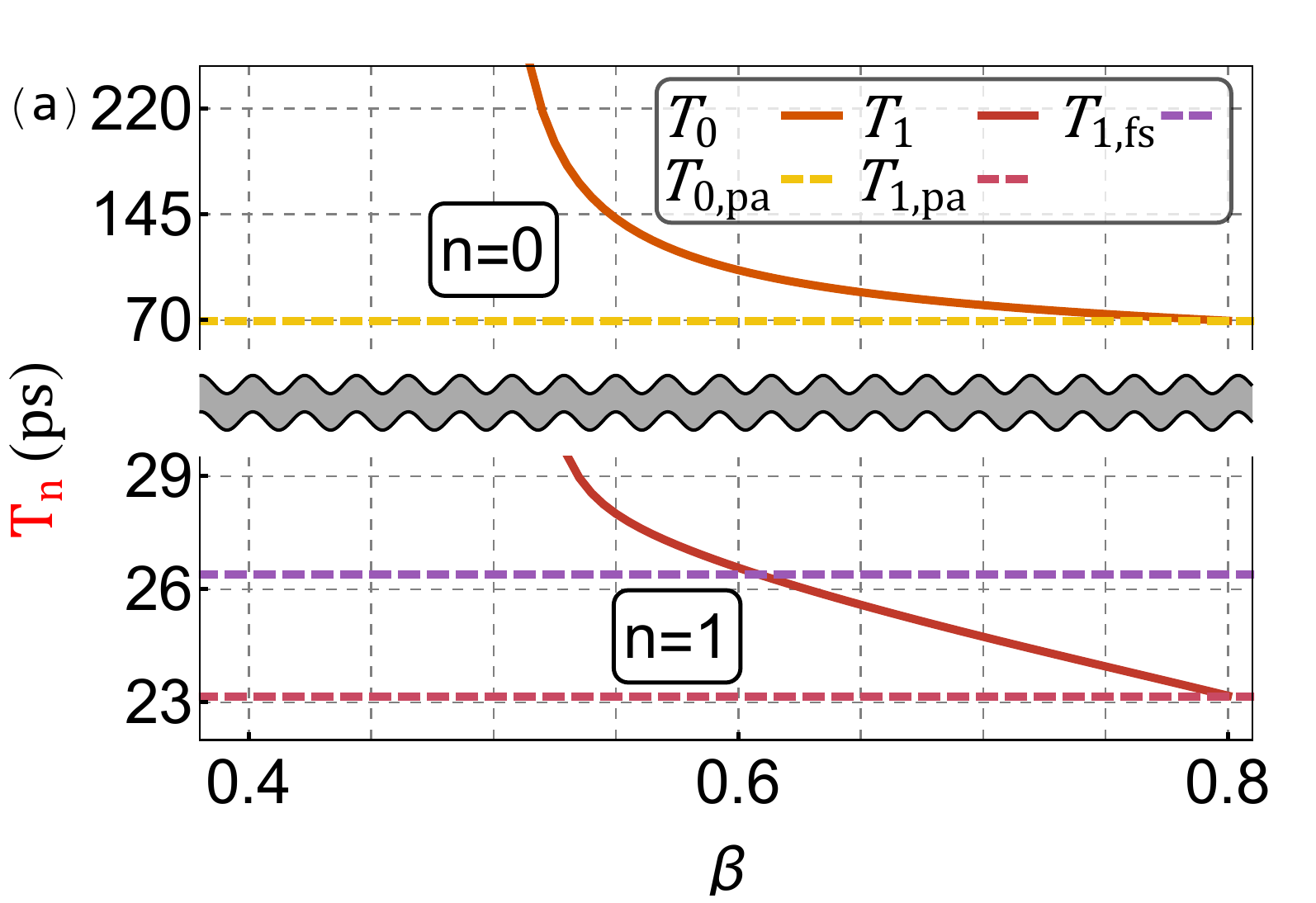}
	\end{minipage}
	\hfill
	\begin{minipage}[b]{\columnwidth}
		\includegraphics[width=\columnwidth]{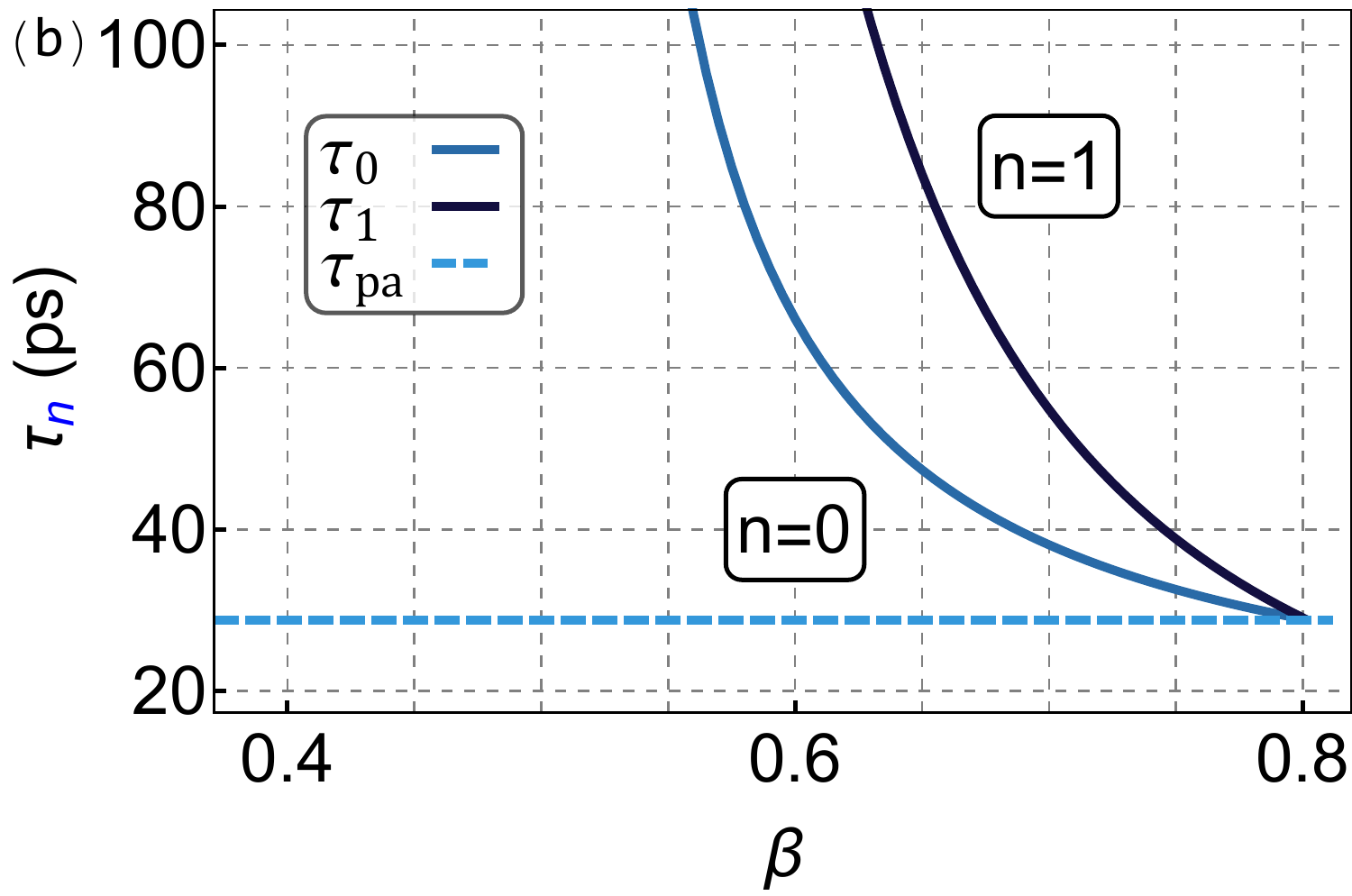}
	\end{minipage}
	\caption{
		(a) period $T_n$ and (b) decay time $\tau_n$ versus $\beta$ for the Ag nanogranular film with $d$=12 nm and $h$=50 nm. The first two modes $n=0,1$ are addressed. EMA model (full lines) and limit cases (dashed lines) obtained for the $fs$ and $pa$ scenarios, respectively. The y-axis in the graph of panel (a) is broken for sake of graphical clarity. The scales above and below the brake are different for ease of representation. $T_n$ diverges for $\beta\rightarrow$0.5, an artefact ascribable to the pitfalls of Budiansky formulas. The $fs$ scenario yields, for mode $n=0$, an infinite period (corresponding to a film translation), hence it is not reported in panel (a). The $fs$ scenario yields an infinite decay time, hence it is not reported in panel (b).}
	\label{fig:limit_case_ema_tot}
\end{figure*}

Following the same procedure, we now perform FEM simulations on the 3D pillar model varying the pillar layer filling fraction, $\alpha$, for a fixed film thickness, $h$=50 nm, and compare the results against the values obtained from the pillar model. Fig.~\ref{fig:per_deca_quality_n_1} reports the oscillation period (a), decay time (b) and quality factor (c), $Q_{n}$=$\pi\left(\tau_{n}/T_{n}\right)$, calculated for $n$=0, where the superscripts $FEM$ and $pil$ stand for FEM simulations and pillar model, respectively. The same quantities, calculated for $n$=1, are reported throughout panels (d-f). The two models yield the same results, the relative differences, $(X^{FEM}_{n}-X^{pil}_{n})/X^{FEM}_{n}$ with $X$=$\{T, \tau, Q\}$,  amounting, at most, to a few percents, see insets to each graph. For the case $n$=0 we were able to perform FEM simulations down to $\alpha$=0.05, whereas for $n$=1 numerical problems impeded extending simulations below $\alpha$=0.3. Given the boosting of $Q_{1}$ for the latter case, FEM simulations, as the one reported in Fig.~\ref{fig:fem_simu} (b), were performed extending the time range to 500 ps. The overall result is that the analytic 1D pillar model perfectly reproduces the results of the more involved 3D FEM pillar model. Furthermore, the former does not pose any problem for the case of low $\alpha$ values, it is order of magnitudes more efficient in terms of computation times (four orders of magnitudes for the present geometry), and, being analytic, it is much more amenable to fit experimental data and clearly identifies the structural parameters leading the acoustic problem.

For low $\alpha$ values simulations were also performed varying the pillar position within the unit cell $xy$ plane and the pillar cross-sectional geometry (a square instead of a circle while keeping the same surface area), the results remaining unaltered. Those of low $\alpha$ values constitute the worse case scenarios for these tests since a slender pillar can be substantially displaced within the unit cell, whereas only small translations can be tested for the case of plumped pillars. i.e. greater $\alpha$ values. These evidences suggest that results are invariant with respect to the specific disordered interface realization. Specifically, the detailed knowledge of the stresses distribution across the interface does not affect the solution in terms of quasi-mode period and lifetime, the relevant aspect rather being the  integral of the stresses exchanged across the interfaces. The latter supports the physical ansatz implied in the 1D pillar model.

For the sake of completeness, in SI we also report the modulus of the displacement $\left|\{\tilde{\textbf{u}}_{n,50}\left(\textbf{r}\right)\}\right|$, for the first (n=0) and the second (n=1) film breathing modes for $\alpha$=0.05, 0.4 and 0.75. These plots give an idea of the quasi-breathing mode evolution from the quasi-$\textit{free standing}$ to the quasi-$\textit{perfect adhesion}$ scenarios.

\section{EMA model}
In order to display the potential of the pillar model, its broad validity range and its added value with respect to more traditional approaches, a simpler 1D model, addressed as Effective Medium Approximation model (EMA) and based on an homogenized interface layer, is now introduced and its dispersion relation calculated. Its limit of validity, restrained to small porosities, are discussed at the light of the pillar model, showing the need for the latter to correctly access the acoustic to structure relation in granular ultra-thin films.
The interface layer, previously identified with the pillar layer, is now accounted for via a continuum, isotropic and homogeneous slab, addressed as $\textit{effective}$ interface layer, see Fig.~\ref{fig:geometry_scheme_ema}. The latter mimics an interface granular layer of thickness $d$, with its solid component made of the same material constituting the NPs and of filling fraction $\beta$. The parameters $d$ and $\beta$ play a similar role as $h$ and $\alpha$ in the pillar model. The elastic properties of the $\textit{effective}$ interface layer, denoted with an asterisk as a superscript, are calculated on the basis of Budiansky theory \cite{budiansky1965elastic}. The bulk, $K^{*} (\beta)$, and shear modulus, $G^{*} (\beta)$, are obtained through:
\begin{align}
\sum\limits_{i=1}^{n} \dfrac{c_{i}}{1 + A\left(\dfrac{K_{i}}{K^{*}}-1\right)}=1 \, ,
\notag
\\
\sum\limits_{i=1}^{n} \dfrac{c_{i}}{1 + B\left(\dfrac{G_{i}}{G^{*}}-1\right)}=1 \, ,
\label{eq:boudiansky}
\end{align}
where the value of $A$ and $B$ are
\begin{equation}
A=\dfrac{1+\nu^{*} (\beta)}{3\left(1-\nu^{*} (\beta)\right)} \, ,
\qquad
B=\dfrac{2\left(4-5\nu^{*} (\beta)\right)}{15\left(1-\nu^{*} (\beta)\right)} \, ,
\label{eq:boudiansky_coef}
\end{equation}
in which the Poisson's ratio is expressed via $K^{*} (\beta)$ and $G^{*} (\beta)$) by the standard relation
\begin{equation}
\nu^{*}=\dfrac{3 K^{*} (\beta) - 2 G^{*} (\beta)}{6 K^{*} (\beta) + 2 G^{*} (\beta)}\, .
\label{eq:boudiansky_Poisson}
\end{equation}
In Eq.(\ref{eq:boudiansky}) $c_i$, $K_{i}$, and $G_{i}$ are the volume fraction, the bulk modulus and the shear modulus of the phase $i$, respectively, where, in the present case, $N=2$, $i=1$ stands for vacuum and $i=2$ for the material constituting the NPS (bulk silver in the following), i.e. $c_2$=$\beta$.
A major pitfall of Budiansky formulas stands in the fact the elastic coefficients vanish when $\beta$ reaches 0.5, thus setting a limit to the applicability of the EMA model, as will be discussed shortly.

Since the transversal contraction is prevented in the $\textit{effective}$ interface layer, the P-wave velocity is $v^{*}(\beta) = \sqrt{\frac{C_{11}^{*} (\beta)}{\rho^{*} (\beta)}}$.

The interface boundary conditions for the EMA model are the ``perfect adhesion'' ones. In the following we summarize the full set of boundary conditions for the EMA model:
\begin{enumerate}
\item free standing at the top of the NP-layer ($z=h$):
\begin{equation}
C_{11}^{NP} \dfrac{\partial u_z^{NP} \left(h,t\right)}{\partial z} = 0 \, ,
\label{eq:BCP1_ema}
\end{equation}
\item continuity of stresses at the interface between the NP-layer and the $\textit{effective}$ homogeneous layer ($z=d$):
\begin{equation}
C_{11}^{NP} \dfrac{\partial u_z^{NP} \left(d,t\right)}{\partial z}  = C_{11}^{*} (\beta)  \dfrac{\partial u_z^{*} \left(d,t\right)}{\partial z} \, ,
\label{eq:BCP2_ema}
\end{equation}
\item continuity of the displacement at the interface between the NP-layer and the $\textit{effective}$ homogeneous layer:
\begin{equation}
u_z^{NP}\left(d,t\right) = u_z^{*}\left(d,t\right) \, ,
\label{eq:BCP3_ema}
\end{equation}
\item continuity of stresses at the interface between the $\textit{effective}$ homogeneous layer and the sapphire substrate ($z=0$):
\begin{equation}
C_{11}^{*} (\beta) \dfrac{\partial u_z^{*} \left(0,t\right)}{\partial z}  = C_{11}^{sub} \dfrac{\partial u_z^{sub} \left(0,t\right)}{\partial z} \, ,
\label{eq:BCP4_ema}
\end{equation}
\item continuity of the displacement condition at the interface between the $\textit{effective}$ homogeneous layer and the sapphire substrate:
\begin{equation}
u_z^{*}\left(0,t\right) = u_z^{sub}\left(0,t\right) \, .
\label{eq:BCP5_ema}
\end{equation}
\end{enumerate}
Enforcing the boundary conditions Eqs.(\ref{eq:BCP1_ema})-(\ref{eq:BCP2_ema})-(\ref{eq:BCP3_ema})-(\ref{eq:BCP4_ema})-(\ref{eq:BCP5_ema}) to Eqs.(\ref{eq:variable_decomposition}) yields the following equation in the unknown $\omega(d,\beta)$:
\begin{widetext}
\begin{equation}
Z^{NP} -
\dfrac{C_{11}^{*} (\beta ) \cot \left(\dfrac{
(h-d) \omega}{v_{z}^{NP}}\right) \left[ v_{z}^{*}(\beta) Z^{sub} \cos \left(\dfrac{d~\omega
}{v_{z}^{*} (\beta)}\right)-i
C_{11}^{*} (\beta ) \sin \left(\dfrac{d~\omega
}{v_{z}^{*} (\beta)}\right)\right]}{v_{z}^{*} (\beta)
\left[ v_{z}^{*} (\beta) Z^{sub} \sin
\left(\dfrac{d~\omega }{v_{z}^{*} (\beta)}\right)+i C_{11}^{*} (\beta ) \cos \left(\dfrac{d~\omega }{v_{z}^{*} (\beta)}\right)\right]}
= 0 \, .
\label{eq:impli_solu_ema}
\end{equation}
\end{widetext}

\noindent
Mutatis mutandis from the pillar mode case, Eq.(\ref{eq:impli_solu_ema}) may be solved numerically and yields, for each fixed set of parameters $\left(d,\beta\right)$, infinitely many complexed-value solutions  $\omega=\omega_n\left(d,\beta\right)$, with $n$=$\{$0,1,2,...$\}$ the index numbering the mode.

The total thickness of the NP-layer assigned, the free parameter in Eq.(\ref{eq:impli_solu_ema}) are the height of the
interface layer, $d$, and its filling fraction, $\beta$. The relations linking the period of vibration, $T_n(d,\beta)$, and the wave decay time, $\tau_n(d,\beta)$, to the n-mode complex-valued angular frequency are expressed through Eqs.(\ref{eq:relation_period_decay}).

Comparison of Eq.(\ref{eq:impli_solu_pil}) and Eq.(\ref{eq:impli_solu_ema}) show that the pillar and EMA models yield the same results provided $d=q$, $\alpha E^{bk}  = C_{11}^{*} \left(\beta \right)$  and $v_z^{pil}=v_z^* \left(\beta\right)$. For the case of Ag NPs, the previous equations are satisfied if $\alpha=\beta=0.770439$.\\

\subsection{The EMA model: Case Study}

\begin{figure*}[t]
\centering
\begin{minipage}[b]{\columnwidth}
\includegraphics[width=\columnwidth]{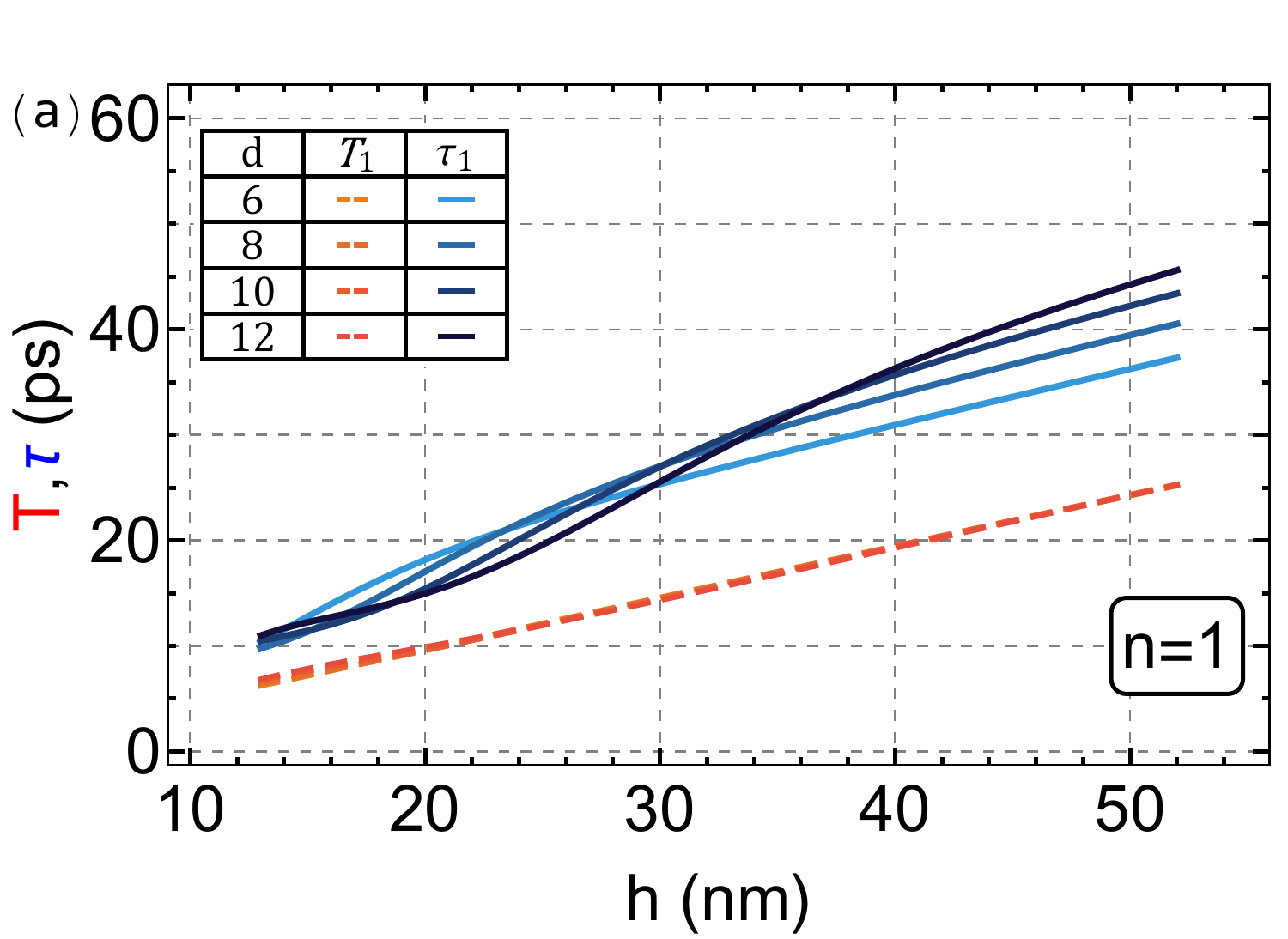}
\end{minipage}
\hfill
\begin{minipage}[b]{\columnwidth}
\includegraphics[width=\columnwidth]{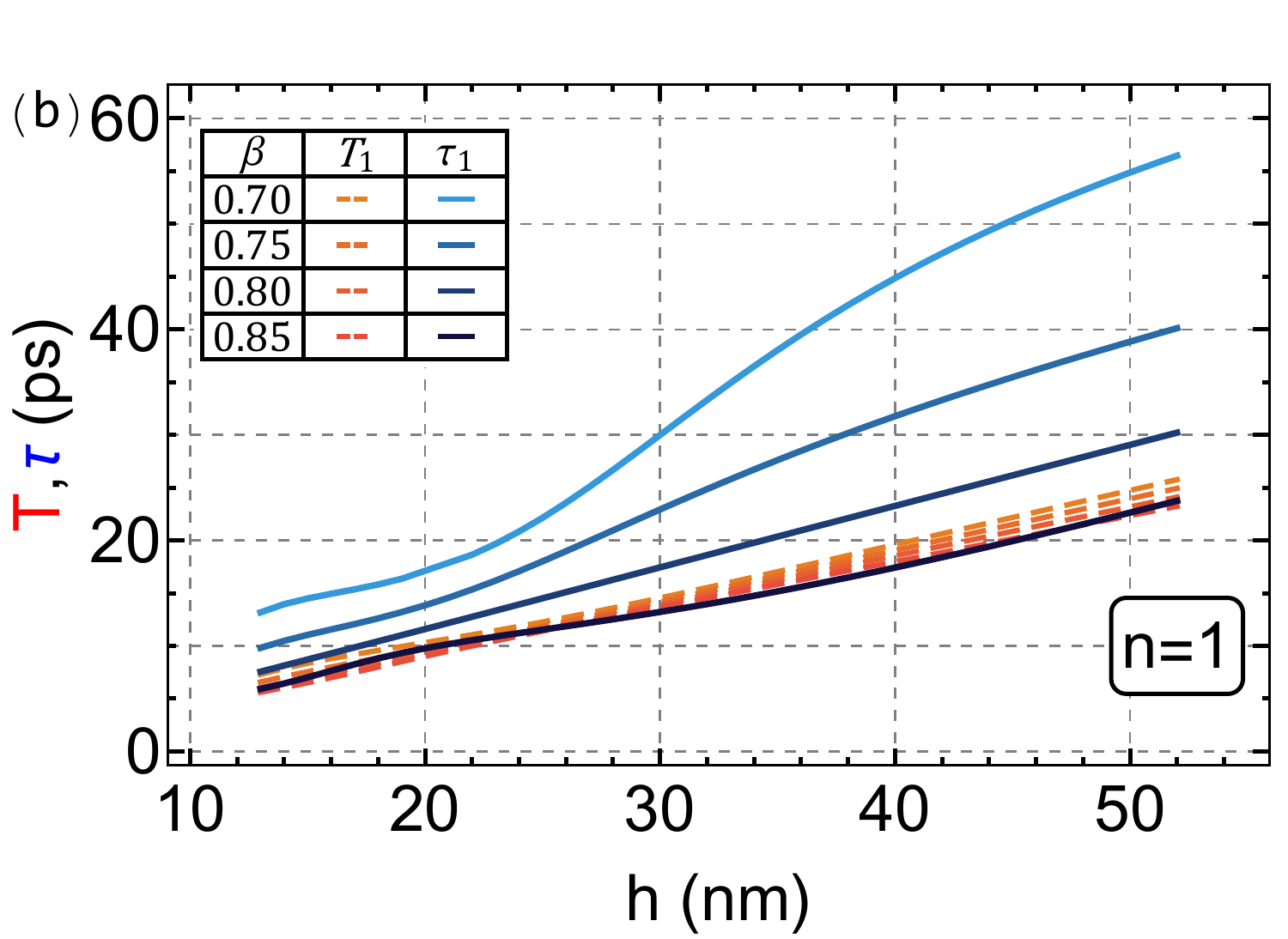}
\end{minipage}
\caption{$T_1(h;d,\beta)$ and $\tau_1(h;d,\beta)$ vs $h$ for $n=1$ for the Ag nanogranular film:
(a) fixed $\beta =  0.73$ while varying $d$ (expressed in nm);
(b) fixed $d = 12$ nm while varying $\beta$.
}
\label{fig:parametric_study_ema_tot}
\end{figure*}
We now exemplify the EMA model considering the same situation addressed in the case study for the pillar model, the only exception being the replacement of the pillar with the $\textit{effective}$ homogeneous film of $d$=12 nm.

The oscillation period $T_{n}$ and lifetime $\tau_{n}$ for the first two modes of the EMA model, $n$=$\{0,1\}$, are reported versus $\beta$ as full lines in Fig.~\ref{fig:limit_case_ema_tot} panel (a) and (b), respectively. For $\beta=0.8$ the density of the $\textit{effective}$ homogeneous layer matches the density of the NP-layer, the latter being 0.8 that of bulk Ag. We do not consider interface densification, the maximum $\beta$ value is thus once again constrained to 0.8.
$T_n$ diverges as $\beta$ approaches 0.5, this being due to the elastic constants becoming null in the Budiansky formula. For the same reason, $T_1$ is not bound between the values $T_{1,pa}$ and $T_{1,fs}$, as should be the case for a correct model. On the contrary, $T_n$ correctly approaches $T_{1,pa}$=23 ps for $\beta \rightarrow$0.8, that is when the interface layer becomes identical to the NPs layer. As for the lifetime, $\tau_{n}$ diverges as $\beta$ approaches 0.5, again due to the pitfalls of Budiansky formulas.
\subsection{The EMA model: Parametric Study}
We here repeat the same parametric study, previously performed for the pillar model, for the case of the EMA model. $T_1(h;d,\beta)$ and $\tau_1(h;d,\beta)$ are reported versus the total thickness $h$ of the NP-layer for a fixed value of $\beta = 0.73$ (the value that gives optimal fitting of the photoacoustic data, see SI) while varying the parameter $d$ across the set of values $\{6,8,10, 12\}$nm, see Fig.~\ref{fig:parametric_study_ema_tot}(a), and vice versa, fixing a value $d=12$ nm (the value that gives optimal fitting of the photoacoustic data, see SI) and varying $\beta$ across the set of values $\{0.70,0.75,0.80, 0.85\}$, see Fig.~\ref{fig:parametric_study_ema_tot}(b). Fig.~\ref{fig:parametric_study_ema_tot} shows the same salient features observed for the pillar model in Fig.~\ref{fig:parametric_study_pil_tot}: the position of the inflection point and the magnitude of the tangent in such point being governed quite independently by $d$ and $\beta$, respectively.

\subsection{Pillar vs EMA model}
\begin{figure}
\centering
\includegraphics[width=\columnwidth]{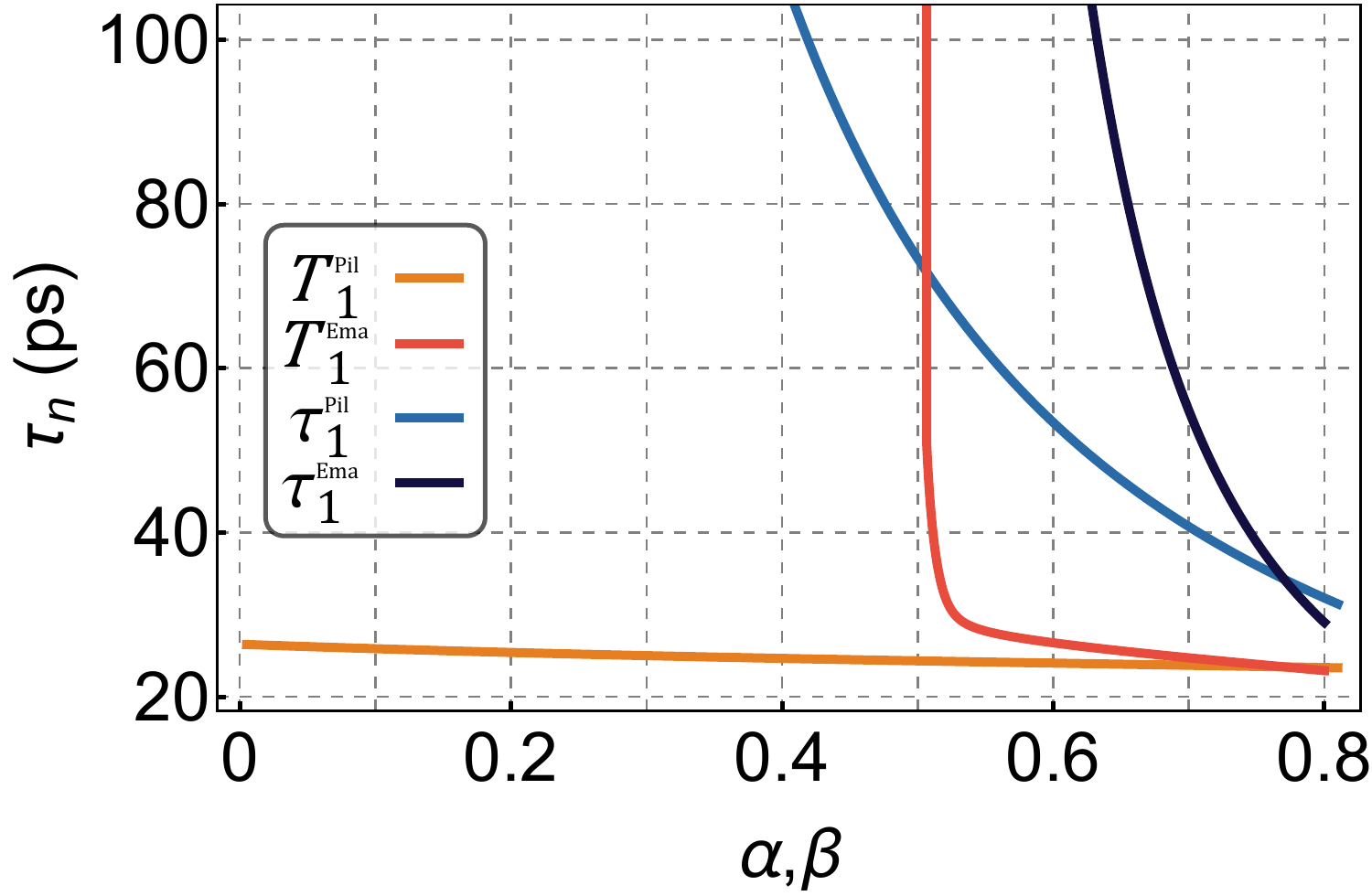}
\caption{Ag nanogranular film with $q$=12 nm (pillar), $d$=12 nm (EMA) and $h$=50 nm.
Decay time $\tau^{pil}_1$ versus $\alpha$ (pillar model) and $\tau^{EMA}_1$ vs $\beta$ (EMA model). The two models coincide for $\alpha=\beta\approx0.77$.}
\label{fig:optimal_solution_ema}
\end{figure}

The pillar model is more adherent to physical reality than the EMA model and, contrary to the latter, is reliable across the entire spectrum of interface filling factor values. The EMA model suffers a major drawback in that both the oscillation periods and decay times diverge as the interface layer filling factor approaches 0.5. The EMA and pillar models yields the same results for a very specific value of the layer filling fraction, which happens to be $\sim$ 0.77 for the case here investigated. The EMA model yields reasonable predictions for small departures of $\beta$ from this values and, in this range, its control parameters work alike the ones of the pillar model. For greater departures of $\beta$  from the optimal value the EMA model fails. Fig.~\ref{fig:optimal_solution_ema}, well summarises these points, reporting, on the same graph and for the same sample, $T^{pil}_1$ and $\tau^{pil}_1$ versus $\alpha$, for the pillar model, and $T^{EMA}_1$ and $\tau^{EMA}_1$ vs $\beta$ for the EMA model.

\section{Conclusions and perspectives}
The pillar model, a fully analytical 1D acoustic model for nanoporous thin films adhered on a flat substrate, was here proposed. The analytical dispersion relation for the frequencies and lifetimes of the film's acoustic breathing modes were obtained in terms of the interface layer's porosity and thickness. The model was successfully benchmarked both against full 3D FEM simulations of a 3D pillar model and photoacoustic data available from the literature on a archetypal model system. The interface mechanical properties of the experimental model system itself bear great applicative relevance, as outlined in recent literature. In order to asses the potential of the pillar model and its broad validity range, its performance was compared against a simpler 1D analytical model, addressed as EMA model, based on an homogenized interface layer of Budiansky type. The limits of applicability of the EMA model were addressed, together with the necessity of deploying the pillar model for most filling factors.

The results here reported are relevant both under a fundamental and applicative stand point. As for the former, the pillar model provides a vivid physical representation of the acoustic of porous thin films and its controlling parameters. More generally, it may be deployed to access the acoustic to structure relation in materials affected by disordered interfaces. The model showed that the physics is primarily dictated by the integral of the stresses exchanged across the interfaces rather than their detailed distribution. Being fully analytical and 1D, the model is computationally very efficient and particularly amenable to fit experimental quasi-breathing mode periods and lifetimes. The model allows accessing the interface layer parameters, which proved challenging to retrieve otherwise. On the other hand, should the porous film morphology been known a-priori, the model correctly predicts its acoustic response. As for applications, knowledge of granular thin films interfaces adhered on a substrate is of paramount importance in a variety of sectors. Just to mention a few, the NP interface layer rules the adhesion properties of bactericidal coatings \cite{benetti2020antimicrobial, benetti2017direct}, both the mechanical and the electrical endurance of bendable transparent conductive oxides \cite{torrisi2019ag} and conductive NPs films produced by inkjet techniques \cite{kao2011} and the sensitivity of photoacoustics sensors \cite{benetti2018photoacoustic}.

The pillar model is scale-invariant and may thus be deployed to investigate systems of greater dimensions, ranging from porous foams for vibration transmission control, to rock sediments laying on a continuous bed to seismological scenarios \cite{peng2021}. Furthermore, the model in applicable, beyond the case of granular materials, to any patched interface. This is the case, for instance, when acoustically addressing the wrinkled interface that may arise between a 2D or a few layers material and its supporting substrate \cite{vialla2020time}, when investigating the acoustic properties of thin films suspended on pillars \cite{chaste2018}, or when inspecting for the presence of PMMA residues between a nano-patterned structure, fabricated via e-beam, lithography and the substrate it adheres on, an issue of the utmost importance in post-processing quality control.

The pillar model also provides a connection to the adhesion forces. Even though a direct comparison with the pull-off force, as provided by the most common adhesion models (JKR\cite{johnson1971surface}, DMT\cite{derjaguin1975effect}), is not straightforward, a simplified average pull-off pressure estimate is presented in SI.

\vspace{0.25cm}
\renewcommand{\arraystretch}{1.3}
\begin{table}[h!]
	\centering
	\begin{tabular}{lrl}
		\hline
		\hline
		$\rho^{NP}$	& 8400 & kg m$^{-3}$ \\[1ex]
		$v_{z}^{NP}$	& 2880 & m s$^{-1}$ \\[1ex]
		$Z^{NP}$	& 2.42$\times$10$^{7}$ & kg s$^{-1}$ m$^{-2}$ \\[1ex]
		$C_{11}^{NP}$	& 6.96$\times$10$^{10}$ & Pa \\[1ex]
		$\rho^{bk}$	& 10490 & kg m$^{-3}$ \\[1ex]
		$v_{z}^{bk}$ & 2740 & m s$^{-1}$ \\[1ex]
		$Z^{bk}$ & 2.87$\times$10$^{7}$ & kg s$^{-1}$ m$^{-2}$ \\[1ex]
		$E^{bk}$	& 7.88$\times$10$^{10}$ & Pa \\[1ex]
		$\rho^{sub}$	& 3986 & kg m$^{-3}$ \\[1ex]
		$v_{z}^{sub}$	& 11260 & m s$^{-1}$ \\[1ex]
		$Z^{sub}$	& 4.49$\times$10$^{7}$ & kg s$^{-1}$ m$^{-2}$ \\[1ex]
		\hline
		\hline
	\end{tabular}
	\caption{Summary of the mechanical properties of the layers}
	\label{tab:para_peli}
\end{table}
\renewcommand{\arraystretch}{1}

\noindent
\textbf{SUPPLEMENTARY INFORMATION}
\\
EMA model best fit solution for mode n=1, displacement field modulus for the first (n=0) and the second (n=1) film breathing modes for several $\alpha$ values, parametric study for the pillar model (fixed $q$ and low values of $\alpha$), computation of the surface energy and pull-off pressure.\\
\\
\noindent
\textbf{AUTHOR INFORMATION.}

\noindent
\textit{Corresponding author}
\\
*\textbf{Giulio Benetti} (giulio.benetti@aovr.veneto.it)
\\
\textit{ORCID}
\\
\textbf{Gianluca Rizzi}: 0000-0002-5967-5403
\\
\textbf{Giulio Benetti}: 0000-0002-7070-0083
\\
\textbf{Claudio Giannetti}: 0000-0003-2664-9492
\\
\textbf{Luca Gavioli}: 0000-0003-2782-7414
\\
\textbf{Francesco Banfi}: 0000-0002-7465-8417
\section*{acknowledgement}
All the authors are grateful to Prof. Bigoni for enlightening discussions regarding the pillar model.
G.R. acknowledge funding from the French Research Agency ANR, METASMART (ANR-17CE08-0006) and the support from IDEXLYON in the framework of the Programme Investissement d$'$ Avenir (ANR-16-IDEX-0005).
C.G. and L.G. acknowledge support from Universit\`a Cattolica del Sacro Cuore through D.2.2 and D.3.1. C.G. acknowledges financial support from MIUR through the PRIN 2017 program (Prot. 20172H2SC4-005). F.B. acknowledges financial support from Universit\'e de Lyon in the frame of the IDEXLYON Project (ANR-16-IDEX-0005) and from Universit\'e Claude Bernard Lyon 1 through the BQR Accueil EC 2019 grant. L.G. acknowledges support from Universit\'e de Lyon as an Invited Professor. 
The authors thank Michael Cappozzo for graphical support in realizing 3D renderings of the different models.
\vspace*{3cm}

\end{document}